\documentclass[
reprint,
amsmath, amssymb,dvipsnames,
aps,
floatfix
]{revtex4-2}

\usepackage{graphicx}
\usepackage{xcolor}
\usepackage{dcolumn}
\usepackage{bm}
\usepackage{animate} 

\pdfoutput=1 
             
\def\be{\begin{equation}}
\def\ee{\end{equation}}
\def\bea{\begin{eqnarray}}
\def\eea{\end{eqnarray}}

\def\yzero{\smash{\hbox{$y\kern-4pt\raise1pt\hbox{${}^\circ$}$}}}

\def\beq{\begin{equation}}
\def\eeq{\end{equation}}
\def\beqa{\begin{eqnarray}}
\def\eeqa{\end{eqnarray}}

\def\-{\hphantom{-}}

\def\s2{\frac{1}{\sqrt2}}

\def\beq{\begin{equation}}
\def\eeq{\end{equation}}
\def\beqa{\begin{eqnarray}}
\def\eeqa{\end{eqnarray}}

\def\IF{\relax{\rm I\kern-.18em F}}
\def\II{\relax{\rm I\kern-.18em I}}
\def\IP{\relax{\rm I\kern-.18em P}}
\def\IC{\relax\hbox{\kern.25em$\inbar\kern-.3em{\rm C}$}}
\def\IR{\relax{\rm I\kern-.18em R}}

\def\Dsl{\,\raise.15ex\hbox{/}\mkern-13.5mu D} 
\def\IZ{Z\kern-.4em  Z}

\begin{document}

\preprint{APS/123-QED}

\title{\boldmath{Q-ball-like solitons on the M2-brane with worldvolume fluxes}}

\author{P. Garcia}
\email{pgarcial@ucab.edu.ve}
\affiliation{Departamento de Física, Facultad de Ingeniería, Universidad Católica Andrés Bello, Caracas 1020-A, Venezuela.} 

\author{M.P. Garcia del Moral}
\email{m-pilar.garciam@unirioja.es}
\affiliation{Área de Física, Departamento de Química, Universidad de la Rioja, La Rioja 26006, Spain.}

\author{J.M. Peña}
\email{joselen@yahoo.com; joselen.pena@ce.ucn.cl}
\affiliation{Departamento de Física, Universidad Católica del Norte, Casa Central. Angamos 0610, Antofagasta, Chile.}

\author{R. Prado-Fuentes\footnote{The order of the authors is alphabetical}}
\email{reginaldo.prado@ua.cl; reginaldo.prado@umayor.cl}
\affiliation{Núcleo de Matemática, Física y Estadística, Facultad de Ciencias, Universidad Mayor, chile.}
\affiliation{Departamento de Física, Universidad de Antofagasta, Aptdo 02800, Chile.}

\date{\today}

\begin{abstract}
In this paper we obtain a family of analytic solutions to the nonlinear partial differential equations that describe the dynamics of the bosonic part of the mass operator of a  M2-brane compactified on $M_9\times T^2$ in the LCG with worldvolume fluxes. Those fluxes can be induced by a constant and quantized supergravity 3-form. This sector of the theory, at supersymmetric level, has the interesting property of having a discrete spectrum. We have focused on the characterization of Q-ball-like (QBL) solitons on the M2-brane with worldvolume fluxes.  Two scenarios are analysed: one in which the system is isotropic and the other anisotropic. In the isotropic case, we obtain analytic families of string-like solutions to the membrane equations of motion in the presence of a non-vanishing symplectic gauge field that satisfy all constraints. We explicitly show a localised family of QBL solutions.  It is demonstrated that although the solutions generally exhibit dispersion, they also allow for dispersion-free solutions. In the non-isotropic case, we obtain  full-fledged membrane QBL solutions by numerical methods. We characterize some other properties of the solutions found. The dynamics of the QBL solutions are also encountered. We analyze the Lorentz boosts and Galilean transformations. Since we work in the Light Cone Gauge, the Lorentz transformed solutions are not automatically solutions, rather some extra conditions must be imposed. Only a subset of the solutions remain. We discuss some examples. The QBL solitons of the M2-brane that have been discovered contain an interaction term between the Noether charge of the Q-ball and the topological monopole charge associated with the worldvolume flux. The monopole charge increases the stability of the analytic solutions against fission. The superposition law of the analytic solutions was obtained, which allows linear superposition, but only for very limited discrete frequencies. Further research is required to determine whether these solutions can be related to a new type of soliton, the Q-monopole ball.

\end{abstract}



\maketitle
\flushbottom

\newpage

\section{Introduction}  Solitons are stable classical solutions with finite energy to nonlinear equations that are able to mimick the behaviour of particles as well as excitations. They are classified into topological and nontopological ones. Topological solitons like monopoles, vortices, instantons and kinks have been extensively studied in the literature. Non-topological solitons have also received a lot of interest for its potential to explain dark matter among others. One simple case corresponds to Q-balls. Q-balls \cite{COLEMAN1985263} are the generic name given to  solitons, originally with spherical topology, which are invariant under a $U(1)$ global symmetry. This symmetry implies the existence of a Noether charge which provides stability against perturbations. Thus global symmetry has been associated to quantum invariants as lepton or barion number. There exists also gauged Q-balls \cite{Q-ballUgauged,MappinggaugedQ-balls} in the case in which the complex scalar fields that parametrized the Q-ball are also charged under a gauge potential $U(1)$. 
There are also other works in which the authors have considered more general global or gauged groups \cite{Loginov2007, ALONSOIZQUIERDO2023}. Dynamics of Q-balls have been studied in \cite{Floratos:2000-Qballs, DynamicsQ-ballsExpandingUniverse} and for the gauged case \cite{Q-ballUgauged, Gulamov_2015}.  Their stability has also been discussed in several papers, see for example \cite{ThePhysicsofQballsThesis}.
Recently, it has been discovered  that it is possible to obtain solitonic solutions that simultaneously share the two types of properties. These type of solitons have been denoted as Q-monopole-ball \cite{Bai2022}.

Solitons with other topologies different from the sphere have been described to less extent  \cite{ALONSOIZQUIERDO2022}. For example, Q-torus, which corresponds to Q-balls with toroidal topology have been also studied in \cite{Q-TorusInN2SupersimetricQED}.
Q-balls have also been used to model dark matter \cite{art:Supersymmetric_Qballs_as_dark_matterKusenko:1997si}, in the context of string theory \cite{FQuevedo:ModuliStarKrippendorf2018}. This paper describes Q-ball solitons from the open and closed string moduli as well as other types of solitons. However, there are very few results of solitons have been reported at the M-theory level.

Classical solutions of membrane theory can be useful to better understand the M-theory properties. First examples of classical solutions to the membrane theory firstly appeared in \cite{nicolaihoppe}. In the context of matrix models, it was discussed in \cite{Hoppe:1997SolutionsMatrixModelEquations}. The existence of membrane solitonic solutions has been discussed in several works. For example in terms of stringy solitary waves in \cite{Restuccia:1998_MembraneSolitons},  in the context of Q-ball matrix model \cite{Soo-Jong_Rey}, as instantonic solutions \cite{FLORATOS:1998_InstantonSolutionsSelf-dualMembrane}, or in terms of membranes formulated on hyperkähler backgrounds \cite{Bergshoeff:1999_SolitonsOnSupermembrane,Portugues_2004}. There are bounds for the experimental detection of Q-balls \cite{ThePhysicsofQballsThesis},
though they are model dependent.

This paper presents analytical and numerical methods to obtain solitonic solutions of the M2-brane with worldvolume fluxes, specifically Q-ball-like solutions. By Q-ball-like (QBL) solutions we refer to solitonic excitations of the membrane that satisfy the characteristic QBL ansatz. We consider a supermembrane on a $M_9\times T^2$ target space with a $C_{\pm}$ flux background. This configuration possesses a magnetic $U(1)$ monopole in the compact sector \cite{mpdm2019:flux, mpgm10Monodromy} associated with the flux content that induces the presence of a magnetic monopole over the worldvolume \cite{MARTINRestucciaTorrealba1998:StabilityM2Compactified}. We search solutions to the nonlinear equations of the M2-brane Hamiltonian on this particular background focusing on the existence of QBL solitons. These solutions are described by complex scalars that model the dynamics of the excitations on the membrane with fluxes, in the non-compact transverse space to the lightcone coordinates. The Hamiltonian that describes the dynamics of the M2-brane contains on top of nontrivial topological terms, non-linear cuartic scalar contributions and a symplectic field strength whose dynamics under certain assumptions becomes coupled. We analyze two different cases: Isotropic analytical solutions that generate string-like configurations with a non-vanishing cost in energy and numerical non-isotropical solutions for the membrane full-fledged configurations. \newline
The string-like configurations, unlike with the string-like spikes associated with the case of a membrane without central charge, carry minimal energy associated with the M2-brane with worldvolume fluxes and the mass terms and thus have dynamics. 
These analytical solutions of the isotropic case are found by solving the equations of motion (EOM) and the constraints of the theory: the area preserving first class constraint and the flux constraint. The flux condition induces a topological central charge restriction.
The different families of solutions found satisfy the Q-ball ansatz and include the case of localized solutions. 
With the aid of numerical simulation, we extend those results to the non-isotropic case in the absence of a symplectic gauge field. We then obtain, QBLnuev full-fledged membrane solutions.  \newline

Having established the existence of these solutions, we characterise some of their properties.
We show that the analytic families of solutions considered exhibit  dispersion generically, but they also admit dispersion-free solutions. 

The dynamics of the Q-ball solitons are characterised by studying their Lorentz transformation and their nonrelativistic limit, the Galilean one. As we are in the Light Cone Gauge (LCG), the transformed solutions are not automatically valid and some extra restrictions must be imposed. We obtain it for some of the analytic families of solutions considered. 
 
There are two types of charges for the solutions found: a non-topological charge related to the Noether charge of the Q-ball and a topological charge related to the existence of a monopole charge resulting from the pull-back of the target $C_{\pm}$ fluxes. The monopole charge defined on the world volume is induced by the nontrivial embedding in the compact space, and the Q-ball solitons are related to the non-compact target space. There is a non-trivial interaction between them induced by the covariant derivative terms at the level of the mass operator. This interaction leads to an enhanced stability.

Finally, we describe their superposition law and demonstrate that, as expected, they cannot be linearly superposed. However, they can be linearly composed  for very restricted values of the Fourier modes.

The paper is structured as follows: 
 Section 2 reviews the formulation of the supermembrane toroidally wrapped on $M_9\times T^2$ subject to a $C_{\pm}$ flux condition and the equations of motion to solve them. Sections 3 to 8 present our results. In Section 3, we obtain analitically a family of exact QBL string-like solutions for the isotropic case. The dispersion of solutions in the presence of a symplectic gauge field is characterized. Additionally, localized solutions for the QBL ansatz are demonstrated. Section 4 characterizes a membrane QBL soliton numerically.  Firstly, we compare the numerical analysis of the isotropic case with the previously obtained analytical results by imposing the ansatz of the so-called 'family'. Then, we extend the analysis to the non-isotropic case, taking into account the complete membrane excitations, when the symplectic gauge vanishes. In the non-isotropic case, the nonlinear character of the complex scalar field becomes explicitly manifest. In section 5 the dynamics of the solution by Lorentz and Galilean transformations is discussed. Section 6 demonstrates a relation between topological and non-topological charges of the M2-brane, improving stability. In Sec. 7 we present a superposition law for the QBL family of solutions that was previously found for the isotropic case. Linear superposition of the solutions is only possible when the frequencies are strongly constrained, consistent with the nonlinear nature of the equations. The results obtained are discussed and concluded in section 8.


\section{The M2-brane with 2-form worldvolume $C_{\pm}$ fluxes}

In this work we analyze the existence of solutions to the dynamics of an M2-brane mass operator, focusing in those of Q-ball-type. The bosonic sector  that we analyze, corresponds to the supermembrane with $C$ fluxes introduced in \cite{mpdm2019:flux, mpgm10Monodromy} that induce a  topological condition denoted central charges which is associated with the presence of a magnetic monopole over the worldvolume of the membrane \cite{MARTINRestucciaTorrealba1998:StabilityM2Compactified}. The M2-brane is formulated on  a target space $M_9\times T^2$ in the Light Cone Gauge (LCG) \cite{Ovalle1, Ovalle2}. 
We define the embedding maps  on the non-compact space $X^m(\tau,\sigma,\rho) :\Sigma \to M_9$ with $m=2,...,8$. The maps $X^{r}(\tau,\sigma,\rho) :\Sigma \to T^2$, where $r=9,10$,  describe the embedding on the compactified 2-torus. These maps act as scalars on the M2-brane worldvolume and as vectors on the target space. The coordinates of the supermembrane worldvolume are labeled by $(\tau, \sigma, \rho)$ parametrizing $\Sigma \times \mathbb{R}$, with $\Sigma$ representing a Riemann surface of genus one and $\mathbb{R}$ parametrizing time.  The maps $X^{r}$ satisfy the winding condition
\begin{equation}
    \oint_{\mathcal{C}_s}dX^{r}=2\pi R^{r}m_{r}^{s} \ ,
\end{equation}
where $m_{r}^{s}$ are the winding numbers and $R^{r}$ are the torus radii. 
  
The M2-brane is restricted by a worldvolume flux condition. It corresponds to a topological condition denoted by the authors as \textit{Central Charge condition}(CC) {\cite{MARTINRestucciaTorrealba1998:StabilityM2Compactified}}, satisfying that  $n=det\mathbb{W}$ where $\mathbb{W}$ denotes the winding matrix of an irreducible wrapping of the M2-brane around the compactified 2-torus. The central charge condition is associated to the existence of a monopole over the worldvolume which is given by the curvature $\widehat{F}=\epsilon_{rs}d X^{r}_h \wedge d X^{s}_h$ associated to a nontrivial $U(1)$ gauge field $\widehat{F}=d\widehat{A}$  defined on the membrane worldvolume. 
\begin{equation}
\label{cargaCentral}
\int_{\Sigma} \widehat{F}=\int_{\Sigma} \epsilon_{rs}d X^{r} \wedge d X^{s}= n A_{T^2}, \quad n \in \mathbb{Z} /\{0\}\ , 
\end{equation}
where $A_{T^2}$ denotes the 2-torus target space area.

The worldvolume flux can be also induced by the pullback of a target-space 2-form flux condition generated by the quantization of the supergravity 3-form in the LCG once that the $X^-$ has been eliminated from the formulation, see, for example,  in \cite{mpdm2019:flux, mpgm10Monodromy}. 
\begin{equation}
\label{fluxcondition}
\int_{T^2} F_2 =  A_{T^2} n. \quad\textrm{with} \quad \in \mathbb{Z} /\{0\}.
\end{equation}
The fluxes of the Hamiltonian can be induced by the presence of a constant quantized 3-form $C_{\pm}$ background. The Hamiltonian of a supermembrane with central charges corresponds to the Hamiltonian of the M2-brane with worldvolume fluxes $C_-$. Its formulation is equivalent to the Hamiltonian of the M2-brane with central charge, plus a constant shift given by the pullback of the $C_+$ contribution \cite{mpdm2019:flux,mpgm10Monodromy}
\begin{equation}
    H_{C_{\pm}}^{fluxes}=H_{C_-}+\int_{\Sigma}C_+=H_{CC}+\int_{\Sigma}C_+\ .
\end{equation}
Both cases are related, being duals for the $C_-$ case and the central charge one. They are defined in terms of an integer $n$, so in the rest of the paper we will use the central charge formulation for simplicity in th3e resolution. 

The closed one-forms $dX_r$ can be decomposed by a Hodge decomposition in terms of two harmonic one-forms $d\widehat{X}_r$ and two exact one-forms $dA_r$, such that 
\begin{equation}
d X_{r}(\sigma,\rho,\tau) = d{X}_{rh}(\sigma,\rho) + dA_{r}(\sigma,\rho,\tau) \,,
\end{equation}
The $dX_{rh}$ is a closed one-form defined in terms of the basis of harmonic forms as $d{X}_{rh}=R^{r}m_{r}^{s} d\widehat{X}_s(\sigma,\rho)$. They are not physical degrees of freedom in the theory but encode topological information of the compactified manifold.
The central charge condition guarantees the existence of the two independent harmonic constant one-forms, $d\widehat{X}_s(\sigma,\rho)$.  The central charge condition also implies the existence a new  dynamical degree of freedom  $A_{r}$ whose associated one-form $\mathbb{A}_r=dA_r=\partial_aA_rd\sigma^a$, with $\partial_a=(\partial_{\sigma},\partial_{\rho})$ describes a symplectic gauge field over the M2-brane worldvolume. It transforms as a symplectic connection under symplectomorphisms. $dA_{r}$ are exact one-forms.

The LCG Hamiltonian of the M2-brane theory with central charge corresponds to {\cite{Ovalle1,Ovalle2}}
\begin{equation}
\begin{aligned}
\label{hamiltonianirred}
H_{CC}&=T^{-2/3}\int_\Sigma d^2\sigma \sqrt{W}\Big[\frac{1}{2}\Big(\frac{P_m}{\sqrt{W}}\Big)^2+\frac{1}{2}\Big(\frac{P_{r}}{\sqrt{W}}\Big)^2\Big]\\ 
&+ T^{-2/3}\int_\Sigma d^2\sigma \sqrt{W}\Big[ \frac{T^{2}}{4}\left\{X^m,X^n\right\}^2 + \\ &+ \frac{T^{2}}{2}(\mathcal{D}_{r} X^m)^2+\frac{T^{2}}{4}(\mathcal{F}_{rs})^2 \Big],
\end{aligned}
\end{equation}
where $T$ stands for the membrane tension and $W$ is the determinant of the induced spatial part component of the foliated metric on the membrane; $\left\{ A,B\right\}=\frac{1}{\sqrt{W}}\epsilon^{ab}\partial_a A\partial_b B;\ $ is the symplectic bracket defined in terms of the harmonic one-forms $d\widehat{X}$ of the Riemann surface with $a,b=1,2$ and $\partial_a=(\partial_{\sigma},\partial_{\rho})$ and  $\sqrt{W}=\frac{1}{2}\epsilon^{ab}\epsilon_{rs}\partial_a\widehat{X}^r\partial_b\widehat{X}^s$.
 The scalar fields $X^m,A_{r}$ have respectively, canonical momenta  $P_m$ and $P_{r}$. The symplectic covariant derivative is defined as
  $\mathcal{D}_{r}\bullet= D_{r}\bullet+\{A_{r},\bullet\}$  with $D_{r}\bullet=2\pi m_{r}^u\theta_{uv} R _{r}\frac{\epsilon^{{a}{b}}}{\sqrt{W}}\partial_{{a}}\widehat{X}^v\partial_{{b}}\bullet.$ \,
 The derivative $D_{r}$ is 
 defined in terms of the moduli of the  2-torus,  the harmonic one-forms $d\widehat{X}_{r}$, and a matrix $\theta_{uv}$ with $u,v=9,10$, related to the monodromy associated to its global description in terms of a torus bundle \cite{GMPR2012}. 
Therefore, the Hamiltonian contains a symplectic curvature $\mathcal{F}$ defined by
\begin{equation}
 \mathcal{F}_{rs}= D_{r} A_s -D_s A_{r} +\left\{ A_{r} ,A_s \right\}.  \,
\end{equation}
The constraints of the theory associated to the local Area Preserving Diffeomorphisms (APD) are:
\begin{equation}\label{APD local}
\mathcal{D}_{r} P_r+\left\{X^m,P_m\right\}\approx 0\ ,
\end{equation}
and to the two APD global constraints, 
\begin{equation}\label{APD global}
\oint_{\mathcal{C}_s} \Big(\Big(\frac{P_{r}}{\sqrt{W}}\Big)\partial_a X^{r}+\Big(\frac{P_m}{\sqrt{W}}\Big)\partial_a X^m\Big)d\sigma^a \approx 0\ .
\end{equation}
In the dynamics of the M2-brane we will only be concerned with the local area preserving diffeomorphisms constraint and the flux constraint.

The dynamics of the M2-brane with worldvolume fluxes is determined by the equations of motion (EOM) and their constraints. The Lagrangian density of the theory can be obtained by performing a usual Legendre transformation of the Hamiltonian density defined as \begin{equation}\mathcal{H}_{T}=\sqrt{W}\mathcal{H}_c+\Lambda \phi,\end{equation} where $\Lambda$ represents a Lagrange multiplier and $\phi=-\phi_{APD}$  represents the APD constraint given by equation (\ref{APD local}). It contains the Lagrangian of MIM2, previously obtained in \cite{Bellorin-Restuccia-2003}, plus a constant term associated to the $C_+$ flux contribution, denoted by $K$.  
\begin{equation}
\label{actionsuperM2centralchargecompleta}
\begin{aligned}
S&=\int d\tau d\sigma^2\mathcal{L}_T=\\
&=-\int d\tau d\sigma^2\sqrt{W}
\Big[\frac{1}{2}(D_i{X}_m)^2 +&\frac{1}{4}\left\{X_{m}, X_{n}\right\}^{2}
\\&& +\frac{1}{4}\mathcal{F}_{ij}^{2}\Big] +K.
\end{aligned}
\end{equation}
 The tension has been fixed to $T=1$.  Since the $K$ contribution gives a non dynamical constant term, added to the action, its equations of motion are equal to those of the M2-brane with central charges. 
The symplectic field strength, and the symplectic covariant derivative now run over the indices $i,j=0,r$, with $r=9,10$.  $A_0=\Lambda$  the lagrange multiplier and $D_0=\partial_{\tau}$. 
As explained in \cite{Hoppe:PhdTesis, Allen-Andersson-Restuccia-2010} the gauge freedom of the system allows to fix $\Lambda=0$ and then the Lagrangian density function $\mathcal{L}=\mathcal{L}_T-K$  reduces to 
\begin{equation}
\label{lagrangeanosuperM2centralchargecompleta}
\begin{aligned}
\sqrt{W}^{-1}\mathcal{L}&=
\frac{1}{2}[(\dot{X}_m)^2 + (\dot{A}_{r})^2]-\frac{1}{4}\left\{X_{m}, X_{n}\right\}^{2}+\\ &- \frac{1}{2}\left(\mathcal{D}_{r} X_m
\right)^2-\frac{1}{4}\mathcal{F}_{rs}^{2}
\end{aligned}
\end{equation}

The nonlinear system of equations that one has to solve is the following:
From  (\ref{lagrangeanosuperM2centralchargecompleta}) we derive the equations of motion for the dynamical fields $X^m(\tau,\sigma,\rho)$ and $A_{r}(\tau,\sigma,\rho)$
\begin{equation}
\label{equationofmotionnoncompact2}
\ddot{X}^m (\tau,\sigma,\rho)= \left\{X_{n}, \left\{X^{n}, X^{m}\right\}\right\}
+\left\{ {X}_{r},\mathcal{D}^{r}{X}^{m}\right\}\ ,
\end{equation}
\begin{equation}
\label{equationofmotionA4}
\begin{aligned}
\ddot{A}_{r}(\tau,\sigma,\rho) = \left\{ \mathcal{D}_{r} X^m, X_{m} \right\}  +\left\{  \mathcal{F}_{rs},{X}^{s} \right\} \ ,
\end{aligned}
\end{equation}
subject to satisfy
the APD constraint of the theory 
\begin{equation}
\label{constraint_A}
\mathcal{D}_{r} \Big(\dot{A}_r\Big)+\left\{X^m,\dot{X}^m\right\}=0\,\quad \text{for}\quad r=9,10,
\end{equation}
and the  worldvolume flux condition (\ref{cargaCentral}), that restricts the winding numbers allowed for the M2-brane. 

In order to find the admissible M2-brane solutions, the system of equations (\ref{equationofmotionnoncompact2}), (\ref{equationofmotionA4}), (\ref{constraint_A}) and (\ref{cargaCentral}) must be solved.  For simplicity we will also assume that one of the scalar fields is constant and we will fix the harmonic sector as in \cite{BRR2005nonperturbative},
\begin{equation} \label{ansatz armonico} X_2=constant, \quad X_{h_{r}}=R_{r}(n_{r}\sigma+m_{r}\rho), \quad r=9,10.\end{equation} 
 For this ansatz $\sqrt{W}=1$. 
 There are some differences with the case analysed in \cite{BRR2005nonperturbative}. The main one is associated with the fact that for the supersymmetric M2-brane, the worldvolume flux condition guarantees the discreteness of the spectrum. Other difference due to the topological restriction imposed is that appearance of new single-valued degrees of freedom, $A_{r}$ that are components  of a symplectic gauge connection $\mathbb{A}=dA_{r}$ in the theory. The EOM for the constant $X_2$ can be expressed in terms of complex variables $Z_a$ with $a=1,2,3$ where $Z_{a}(\tau , \sigma , \rho) = X_{2a+1}+iX_{2a+2}$. They are, 
 
\begin{eqnarray}\label{equationofmotionnoncompactcomplexZ}
\ddot{Z}_c&= \frac{1}{2}\sum_{a=1}^{3}\left(\left\{\left\{Z_{c}, Z_{a}\right\},  \overline{Z}_{a}\right\}+\left\{\left\{Z_{c}, \overline{Z}_{a}\right\},  {Z}_{a}\right\}\right) + \nonumber\\
&+ \sum_{r=9}^{10}\left\{X_{r},\mathcal{D}_{r} Z_c\right\},
\end{eqnarray}
\begin{eqnarray}
\label{equationofmotionAcomplex}
\ddot{A}_{r}&=
\frac{1}{2}\sum_{a=1}^{3}\left(\left\{\mathcal{D}_{r}Z_a,\overline{Z}_{a}\right\} +\left\{\mathcal{D}_{r}\overline{Z}_a,Z_{a}\right\}\right) + \nonumber \\ 
&+\sum_{s=9}^{10}\left\{\mathcal{F}_{rs},X^{s}\right\} \ ,
\end{eqnarray}
subject to the APD constraint of the theory: 
\begin{equation}
\label{constraint_complex_general}
\frac{1}{2}\sum_{c=1}^{3}\left(\left\{\dot{Z}_{c}, \bar{Z}_{c}\right\}+\left\{\dot{\bar{Z}}_{c}, Z_{c}\right\}\right) -\mathcal{D}_{r}\dot{A}_{r}=0 \ .
\end{equation}
%
 
In the following, we assume the Q-ball ansatz for the complex scalars $Z_a$. We will denote this last ansatz as 'Q-ball like' (QBL) irrespective of its  toroidal topology associated to the membrane worldvolume 
\begin{equation}
\label{QBL_ansatz}
Z_a=f_a(\sigma,\rho)e^{i\omega_a\tau},\quad \text{with} \quad f(\sigma,\rho) \in \mathbb{R} \,.
\end{equation}

The EOM  for $Z_a$ become 
{\fontsize{7}{7} 
\begin{equation}
\begin{aligned}
\label{EOM_Zc-Qball}
\omega_c^2 f_c =&
-
\sum_{r=9}^{10}
\left[
\partial_{\sigma}^{2} f_c
\left[R_{r} m_{r}+\partial_{\rho} A_{r}\right]^{2}
+
\partial_{\rho}^{2}f_c 
\left(R_{r} n_{r}+\partial_{\sigma} A_{r}\right)^{2}
\right]
\\
&+\sum_{r=9}^{10}
\partial_{\sigma \rho}^{2} f_c \left[
2 \ \left(R_{r} n_{r}+\partial_{\sigma} A_{r}\right)  \left(R_{r} m_{r}+\partial_{\rho} A_{r}\right)
\right]
\nonumber\\
&+
\sum_{r}\left[
\partial_{\sigma}^{2} A^ r
\partial_{\rho} f_{c}
\left(R_{r} m_{r}  + \partial_\rho A_{r}\right)
+
\partial_{\rho}^{2} A^ r
\partial_{\sigma} f_c
\left( R_{r} n_{r} + \partial_\sigma A_{r}\right)
\right]
\nonumber\\
&-
\sum_{r}
\partial_{\sigma \rho}^{2} A_{r}
\left[
\partial_{\sigma} f_c 
\left( R_{r} m_{r} + \partial_\rho A_{r}\right)
        +
\partial_{\rho} f_{c}
\left(R_{r} n_{r}  + \partial_\sigma A_{r}\right)
\right]
\nonumber
\\
&- 
\sum_{a}               
    \left[
  \partial_{\sigma}^{2} f_c
(\partial_\rho f_a )^2
-2\partial_{\sigma \rho}^{2} f_c
    \left[\partial_{\sigma} f_{a} \partial_{\rho} {f}_{a}\right]
 +\partial_{\rho}^{2}f_c 
\left(\partial_ \sigma f_{a} \right)^2
                \right]
\nonumber
\\
&+ 
\sum_{a}               
    \left[
\partial_{\rho}^{2} f_a 
\partial_{\sigma} f_c \partial_{\sigma} {f}_{a}
-
\partial_{\sigma \rho}^{2} f_{a} 
(\partial_{\sigma} f_c \partial_{\rho} {f}_{a}
+
\partial_{\rho} f_{c}\partial_{\sigma} {f}_{a}
)\right]
\ .
\nonumber \\
&+\sum_{a} 
\partial_{\sigma}^{2} f_a 
\partial_{\rho} f_{c} \partial_{\rho} {f}_{a}
\ .
\nonumber
\end{aligned}
\end{equation}
}

The EOM for $A_{r}$ are:
{\fontsize{7}{7} 
\begin{eqnarray}\label{EOMForArZiagualfg-Qball}
\ddot{A}_{r}
=&-&
\sum_{a=1}^{3}  
\left[
\partial_{\sigma}^{2} f_{a}\  \partial_{\rho} {f}_{a}
\left(R_{r} m_{r}+\partial_{\rho} A_{r}\right)
+
\partial_{\rho}^{2} f_{a}\  \partial_{\sigma} {f}_{a}\left(R_{r} n_{r}+\partial_{\sigma} A_{r}\right)\right] \nonumber
\\
&+&
\sum_{a=1}^{3}  \left[
\partial_{\sigma \rho}^{2} f_{a}\left[\partial_{\rho} {f}_{a}\left(R_{r} n_{r}+\partial_{\sigma} A_{r}\right)+\partial_{\sigma} {f}_{a}\left(R_{r} m_{r}+\partial_{\rho} A_{r}\right)\right] 
\right]
\nonumber
\\
&+&
\sum_{a=1}^{3}\left[
\partial_{\sigma}^{2} A_{r} \left(\partial_{\rho} f_{a}\right)^2
-
2\ \partial^2_{\sigma \rho} A_{r}\ \partial_{\sigma} f_{a} \partial_{\rho} {f}_{a}
+
\partial_{\rho}^{2} A_{r} \left(\partial_{\sigma} f_{a}\right)^2
\right]
\nonumber
\\
&+&
\partial_{\sigma}^{2} A_{r}  
\left(R_{s} m_{s}+\partial_{\rho}A_{s}\right)^{2}
+\partial_{\rho}^{2} A_{r}
    \left(R_{s} n_{s}+\partial_{\sigma} A_{s}\right)^{2}
\nonumber
\\
&-&
\partial^2_{\sigma \rho} A_{r}
\left[
    2 \left(R_{s} n_{s}+\partial_{\sigma} A_{s}\right)\left(R_{s} m_{s}+\partial_{\rho} A_{s}\right)
    \right] 
\nonumber
\\
&+&
\partial_{\sigma\rho}^{2} A_{s}\left(R_{r} n_{r}+\partial_{\sigma} A_{r}\right)\left(R_{s} m_{s}+\partial_{\rho} A_{s}\right)+ \\
&+&\partial_{\sigma\rho}^{2} A_{s}\left(R_{r} m_{r}+\partial_{\rho} A_{r}\right)\left(R_{s} n_{s} + \partial_{\sigma} A_{s}\right)
\nonumber\\
&-&
\partial_{\sigma}^{2} A_{s}\left(R_{r} m_{r}+\partial_{\rho} A_{r}\right)\left(R_{s} m_{s}+\partial_{\rho} A_{s}\right)+ \nonumber\\
&-&
\partial_{\rho}^{2} A_{s}\left(R_{r} n_{r}+\partial_{\sigma} A_{r}\right)\left(R_{s} n_{s}+\partial_{\sigma} A_{s}\right)
\ ,\nonumber
\end{eqnarray}
        }
where the mixed  partial second order derivatives are assumed to satisfy $\partial^2_{\sigma\rho}=\partial^2_{\rho\sigma}$.

The M2-brane is subject to the area preserving diffeomorphism (APD) constraint 
{
\begin{eqnarray}
\label{constraint_QBall}
0&=&
\sum_{r=9}^{10}R_{r}(\partial_\sigma \dot{A}_{r} m_{r} -  \partial_\rho \dot{A}_{r} n_{r})+
\\&&+\sum_{r=9}^{10}(\partial_\sigma \dot{A}_{r} \partial_\rho A_{r} -  \partial_\rho \dot{A}_{r} \partial_\sigma A_{r}).
\end{eqnarray}}
and subject to the flux condition that restricts the allowed winding numbers appearing in $X_{rh}$,
\begin{equation}\label{central charge}
   n=det(m_{r}^{s})= n_{9}m_{10}-n_{10}m_{9}\ne0 \quad \textrm{with}\quad n\in \mathbb{Z}.
\end{equation}
%
%


\section{Analytic String Q-ball-like (QBL) solutions to the M2-brane with fluxes.} 
In this section we obtain analytic solutions to the system of equations of the M2-brane with $C_{\pm}$ fluxes (\ref{EOM_Zc-Qball})-(\ref{central charge})  of the complex scalar field $Z_a$. In this section, for simplicity, we assume the system to be isotropic $Z_a=Z_b$ for $a\ne b$, having fixed $X_2=0$. This assumption trivialises the symplectic bracket for the complex scalar embedding field and correspomd to strong -like configurations but with cost in energy.  The solutions still contain a non-vanishing symplectic gauge field, 
$\mathbb{A}=dA$ with a non-trivial curvature $\mathcal{F}$. To solve the equations of motion we assume that $A_r$ and $Z$ have the same spatial dependence associated with a single-valued function $f(\sigma, \rho)$. This assumption transforms the equations of motion into a system of three nonlinear coupled differential equations which must also satisfy the APD constraint and the flux constraint, i.e. a non-vanishing central charge.The configurations obtained depend non-trivially on a combination of the spatial variables. In the first two families we illustrate the simplest solutions, in which the dependence on the symplectic gauge cuvature becomes generalized and in the last family we show the existence of localized excitations with the characteristic breathing modes.

Interestingly, due to the central charge condition mass terms are generated, these dominate in the quartic potential along the flat directions and resemble a  modified harmonic oscillator. They correspond to the minimum energy of the theory. 
These contributions are not present in compactifications without induced 2-form worldvolume fluxes. Furthermore, for the solutions obtained, the central charge condition is responsible for generating terms in the equations of motion that allow the presence of breathing modes different from zero.\newline

In order to find analytic solutions to the M2-brane equations, we have inspected the symmetries of the theory. First of all, it is easy to see that there is a symmetry in the choice of the spatial coordinates. Furthermore, the equations are non-linear with triple and quadratic products in the partial derivatives of first and second order.  By direct observation of the differential equations one can observe that a family of solutions for a given $f$ appears when first derivatives are proportionals among them and consequently the same type of relation holds for the second derivatives. Then, a natural choice is to convert the PDE of the system into an ODE-type system by means of imposing that $\partial_{\sigma}f\propto \partial_{\rho}f$-. We keep the partial derivative notation, to recall that the system truly depends on two spatial variables. We can then extend these relations into second order derivatives, then
we will look for a family of ansätze that satisfy the following relations.
\begin{eqnarray}\label{FamiliaGeneral}
\partial_\sigma f_a=j_a\partial_\rho f_a  \quad \textrm{with}\quad a=1,2,3.
\end{eqnarray}

The membrane is closed and therefore all single-valued functions must be periodic on the coordinates ($\sigma,\rho$)  of the torus. For simplicity we are assuming a rectangular 2-torus with the moduli described by the two radii ($R_9, R_{10}$). 
For the $A_r(\sigma,\rho, \tau)$ scalar field, we impose the same type of dependence $\partial_\sigma A_r\propto \partial_\rho A_r$ at the level of the first derivatives. We impose the following relation of proportionality 
\begin{eqnarray}\label{FamiliaGeneral-A}
\partial_\sigma A_r=j_r\partial_\rho A_r, \, \quad r,s=9,10.  \end{eqnarray}
The proportionality in the first derivatives imposes the following 
second derivatives relation 

\begin{equation}
\begin{aligned}
 \partial_\sigma^2
f_a = j_a^2\partial_\rho^2f_a,\quad 
\partial_{\sigma\rho}^2
f_a = j_a \partial_\rho^2 f_a;\qquad
\\ \partial_\sigma^2 
A_r = j_r^2\partial_\rho^2A_r,\quad 
\partial_{\sigma\rho}^2
A_r = j_r \partial_\rho^2 A_r \,.   
\end{aligned}
\end{equation}

We can observe that the EOM is greatly simplified under these assumptions.
For $Z_c$,
%
\begin{equation}
\label{EOMZQballfamilia}
\begin{aligned}
\omega_c^2 f_c =&-\sum_{r=9}^{10}\left[
R_r(j_c m_r -n_r)
\right]^{2}
\partial_{\rho}^{2}f_c \\
&-2\sum_{r=9}^{10}
R_r(j_c m_r -n_r)(j_c-j_r)\partial_\rho A_r
\partial_{\rho}^{2}f_c +\\
&-\sum_{r=9}^{10}
(j_c-j_r)^2(\partial_\rho A_r)^2
\partial_{\rho}^{2}f_c +\\
&+\sum_{r}
R_{r} 
\left( m_r j_r -n_r\right)(j_r-j_c)
\partial_{\rho} f_{c}
\partial_{\rho}^{2} A^ r
\\
&-\sum_{a} (j_c-j_a)^2             
\partial_{\rho}^{2} f_c
(\partial_\rho f_a )^2 \ .    
\end{aligned}
\end{equation}

For $A_r$,
%
{
\begin{equation}
\label{EOMArQballfamilia} 
\begin{aligned}
\ddot{A}_r = &\sum_{a=1}^3\left(j_r-j_a\right)^2\left(\partial_\rho f_a\right)^2 \partial_\rho^2 A_r + \\  +&\sum_{s=9}^{10}\partial_\rho^2 A_r\left[R_s\left(j_r m_s-n_s\right)+\left(j_r-j_s\right) \partial_\rho A_r\right]^2  \\
-& \sum_{s=9}^{10}\partial_\rho^2 A_s
R_r\left(j_s m_r-n_r\right)R_s\left(j_s m_s-n_s\right)   
 \\
+& \sum_{s=9}^{10}\partial_\rho^2 A_s
\sum_{s=9}^{10}\left(j_s-j_r\right) \partial_\rho A_r R_s\left(j_s m_s-n_s\right) \,,  
\end{aligned} 
\end{equation}}
where $r\neq s$. The APD constraint (\ref{constraint_QBall}) with the above ansatz (\ref{FamiliaGeneral-A}) becomes:

\begin{eqnarray}
\label{constraint_Ansatz_en_Ar}
0&=&
\sum_{r=9}^{10}R_r(j_{r} m_r - n_r )\partial_\rho \dot A_{r} \,,
\end{eqnarray}
and the topological constraint (\ref{central charge}) remains invariant.  In the following we propose two different forms to obtain a family of solutions. 

\paragraph{First family of analytic solutions} We propose  the symplectic field $\mathbb{A}=dA$ defined on the compact sector of the theory is related with the QBL solution of the noncompact sector, (\ref{QBL_ansatz}) $Z_a=f_a( \sigma,\rho) e^{i\omega \tau}$  through the following ansatz for the real scalar field 
\begin{equation}\label{A1} A_r=a_r \sum_a f_a(\sigma,\rho)+c_r \tau.\end{equation}
We assume the isotropic case $f_a=f, \, \forall a$, and impose the same proportionality constant for partial derivatives of $A_r$ and $f_a$ such that $j_a=j=j_r$ with $a=1,2,3$ and $r=9,10$. The two equations of motion for $A_r$ become again simplified,
\begin{equation}
\begin{aligned}
0 =& 
 a_r\left[R_s\left(j_r m_s-n_s\right)\right]^2  +\\
 &-a_s\left[R_r\left(j_s m_r-n_r\right) R_s\left(j_s m_s-n_s\right) \right] \,.
\end{aligned}
\end{equation}
By analyzing both equations we arrive to the following  relation, 
\begin{equation}\label{eq:CondF2}
a_{10}^2{R_{9}^2\left(j m_{9}-n_{9}\right)^2}=a_{9}^2 {R_{10}^2\left(j m_{10}-n_{10}\right)}^2 \,,
\end{equation} 
which may be used to establish a relation between the amplitudes $a_r$ of the two fields $A_r(\sigma,\rho,\tau)$ given by 
\begin{equation}\label{a}
 a=a_9/a_{10}=\pm\frac{R_9(km_9-ln_9)}{R_{10}(km_{10}-ln_{10})}.
 \end{equation}
The EOM associated with  Z becomes reduced to
\begin{equation}\label{eq:F2Disper}
\omega^{2} f+\partial_{\rho}^{2} f \sum_{r} R_{r}^{2}\left(n_{r}-j\  m_{r}\right)^{2}=0.
\end{equation}
%
It is an eigenvalue-like equation for $f$,
$\label{eq:F2EDReducida}
\partial_{\rho}^{2} f(\sigma,\rho) =-\lambda_\rho^2 f(\sigma,\rho) 
$
with
\begin{equation}    \lambda_\rho^2 =\frac{\omega^{2}}{ \sum_{r} R_{r}^{2}\left(n_{r}-j\  m_{r}\right)^{2}} \,.
\end{equation}

Observe that in order to guarantee the periodicity condition on $f$, then $\lambda_\rho$ must be integer which imposes $\omega$ to be discrete,
\begin{equation}\label{eq:DispercionFamilia}
    \omega^{2} = \lambda_\rho^2 \sum_{r} R_{r}^{2}\left(n_{r}-j\  m_{r}\right)^{2} \,,
\end{equation}
being proportional to the previous solution already identified, whose solution is
\begin{equation}\label{eq9D:SolF2-1}
f(\sigma, \rho)=c_{2}(\sigma) \sin \left( \lambda_\rho\rho \right)+c_{1}(\sigma) \cos \left(\lambda_\rho\rho\right) \,.
\end{equation}
 By considering  
 $\partial_\rho f=j^{-1}\partial_\sigma f$ 
  , derive
\begin{equation}
\partial_{\sigma}^{2} f(\sigma,\rho) =-\lambda_\sigma^2 f(\sigma,\rho) \,,
\end{equation}
and the solution is
\begin{equation}\label{eq9D:SolF2-2}
f(\sigma, \rho)=c_{1}(\rho) \sin \left( \lambda_\sigma \sigma \right)+c_{2}(\rho) \cos \left(\lambda_\sigma\sigma\right)\,. 
\end{equation}
Hence, the relation between the eingenvalue constants is  $\lambda_\sigma = j \lambda_\rho$. From (\ref{eq9D:SolF2-1}) and (\ref{eq9D:SolF2-2}) 
it is possible to obtain two general solutions, 
{\small
\begin{equation}
\begin{aligned}
f_1(\sigma, \rho)=& c_1^1 \sin \left(j\lambda_\rho\sigma \right) \cos\left(\lambda_\rho\rho\right)+ c_2^1 \cos(j\lambda_\rho \sigma) \sin \left(\lambda_\rho\rho\right) 
\\
f_2(\sigma, \rho)= & c_1^2 \cos \left(j\lambda_\rho\sigma \right) \cos\left(\lambda_\rho\rho \right)+c_2^2 \sin(j\lambda_\rho \sigma) \sin \left(\lambda_\rho\rho\right) .
\end{aligned}
\end{equation}
}

In order to find a solution of the complete system
(\ref{FamiliaGeneral}) we fix the proportionality constant $j\in \mathbb{Q}$ such that $j\lambda_\rho\in \mathbb{Z}$. For the case when the arbitrary constants $c_a^b=r_b$ for different $a=1,2$ , the solutions become reduced to
\begin{equation}
\begin{aligned}
f_1(\sigma, \rho)=& r_1 \sin \left(\lambda_\rho(j \sigma+\rho)\right)\\
f_2(\sigma, \rho)=& r_2\cos \left(\lambda_\rho(j \sigma+\rho)\right) \,.
   \end{aligned}
 \end{equation}
and they become string-like solutions to the membrane equations.
For $j=k/l$ with $k,l\in\mathbb{Z}$, and since the problem is symmetric in the worldvolume coordinates $(\sigma,\rho)$,  the eigenvalues of the function $f$ can be re-expressed as $\lambda_\sigma^2=k^2\lambda^2$ 
and $\lambda_\rho^2=l^2 \lambda^2$ with $\lambda$ integer or rational and $r_1$ and $r_2$ arbitrary constants representing the amplitudes
   \begin{equation}
   f(\sigma, \rho)= r_1 \sin \left(\lambda(k \sigma+l \rho)\right)
+
r_2\cos \left(\lambda(k \sigma+l \rho)\right) \,.
\end{equation}
Finally, the string-like solutions to the equations of motion for the M2-brane with the (\ref{A1}) ansatz are :

{\small
\begin{equation}
\begin{aligned}
    Z(\sigma,\rho)=&( r_1\sin(\lambda(k\sigma+l\rho))+ r_2\cos(\lambda(k\sigma+l\rho)))e^{i \omega \tau} \,,
\\
A_r(\sigma,\rho,\tau)=& a_r(r_1\sin(\lambda(k\sigma+l\rho))+ r_2\cos(\lambda(k\sigma+l\rho))+c_r\tau \,,
\end{aligned}
\end{equation}}

when the breathing modes $\omega$ have the following dispersion relation (\ref{eq:DispercionFamilia}), 
\begin{equation}\label{frecuencia omega}
\begin{aligned}
 \omega =\lambda \sqrt{\sum_r R_r^2 (km_r-ln_r)^2},
\end{aligned}
\end{equation} 

and using $l\lambda=k$, and the amplitudes ratio $a$ of $A_r$  determined through equation (\ref{a}). See that for each pair of Fourier modes there is a given relation between the amplitudes associated to the $A_r$ allowed.
See figure 1 that illustrate the behaviour of the $Z_a$ solution.
The APD constraint is identically satisfied and the central charge condition can be also satisfied for a proper choice of the winding numbers. 

\begin{figure}
    \centering
    \begin{tabular}{c}
\animategraphics[loop,autoplay,width=0.23\textwidth]{5}{GiffF1/F1C-}{0}{49} 
 \animategraphics[loop,autoplay,width=0.23\textwidth]{5}{GiffF1/F1T-}{0}{49}
    \end{tabular}
    \caption{Maple animated solutions of the first two families analyzed, representing the behaviour of $Z_a$, plotted 
    on cartesian coordinates and on the 2-torus toroidal coordinates, for the following choice of parameters $k=2,$ $l=3,$ $r_1=r_2=1,$ $ n_1=m_2=1, m_1=n_2=0,$ $ R_1=3, R_2=2,$ $\lambda=1, $ $ \omega =5 $.  Use Adobe Acrobat Reader. for a correct visualization.
    \label{fig:F1Qball}}
\end{figure}
These solutions trivialize the symplectic bracket, however they do not correspond to trivial solutions. They represent the lowest energy mass operator excitations. These type of solutions, in spite of its simplicity are solitonic due to the non-linear behaviour of the system, as we will see.  An important point in order to characterize the solitonic behaviour of these solutions is to analyze the existence of dispersion.  It serves to distinguish linear from nonlinear behaviors and the dynamics of the solutions. Standard solitonic solutions are those in which the dispersion relations are compensated by the nonlinearities of the system, resulting in stable solutions.  It is important to note that the presence of nonlinearity does not necessarily imply any loss of energy. This is the situation that we analyze here. In all of these solutions the energy is preserved. This fact can be understood since we are describing stable excitations of a single membrane propagating along the $M_9\times T^2$ space time. Since the M2-brane worldvolume has two spatial dimensions, these quantities must be defined in terms of their associated velocity vectors.  An interesting property is that the frequency $\omega$ takes discrete values for a given moduli. This is not an obvious property. The membrane worldvolume is compact, however this does not automatically imply that the spectrum will be automatically discrete, a very famous counter-example is the 11D compact supermembrane whose mass operator which has continuous spectrum from $[0,+\infty)$ \cite{dwln}.  In \cite{mpgm8} the authors rigorously prove for the bosonic membrane that the spectrum is discrete from the monopole charge contribution induced by the central charge (C-fluxes) until infinity.
\paragraph{Dispersion} In the following we will see the dispersion relations for the Q-ball-like ansatzs. Following \cite{book:AtmosphericGeoffrey,book:aguero2020Introduccion} 
the group velocity vector is defined as, 

\begin{equation}
    \textbf{v}_g=\frac{\partial\omega}{\partial \textbf{k}} \equiv \left(\frac{\partial\omega}{\partial k},\frac{\partial\omega}{\partial l}\right)=\nabla_\textbf{k} \omega \,.
\end{equation}

For the dispersion relation (\ref{frecuencia omega}), the group velocity vector is
\begin{equation}
   \textbf{v}_g= \frac{{\lambda}^2 }{\omega} \sum_r  R_r^2 \left((k m_r-l n_r\right)
  (m_r,-n_r)
\end{equation}
and its magnitude is given by,
\begin{equation}
\begin{aligned}
   \vert\textbf{v}_g\vert=\frac{{\lambda}^2 }{\omega} [ \, 
   (\sum_r n_r R_r^2 \left(k m_r-l n_r\right))^2 \\
   +(\sum_r m_r R_r^2 \left(k m_r-l n_r\right))^2 \, ] ^{1/2} \,.    
\end{aligned}
\end{equation} 

The phase speed is defined as 
\begin{equation}
    c(\textbf{k})=\frac{\omega}{|\textbf{k}|}, 
\end{equation}
 where $\textbf{k}=\sqrt{k^2+l^2}$.  It is possible to define also the phase speed along each of the directions as, 
$(c^\sigma=\frac{\omega}{k}, c^\rho=\frac{\omega}{l})$. See, that they are not the components of a vector, i.e. $c(\textbf{k})\neq\sqrt{(c_b^\sigma)^2+(c_b^\rho)^2}$. 

%
The phase speed for the dispersion relation (\ref{frecuencia omega})is given by 
\begin{equation}
    c(\textbf{k})=\lambda \frac{\sqrt{\sum_r R_r^2 (km_r-ln_r)^2}}{\sqrt{k^2+l^2}} \,.
\end{equation}

\begin{figure}
\begin{center}
\begin{tabular}{ c }
 \includegraphics[width=0.5\textwidth]{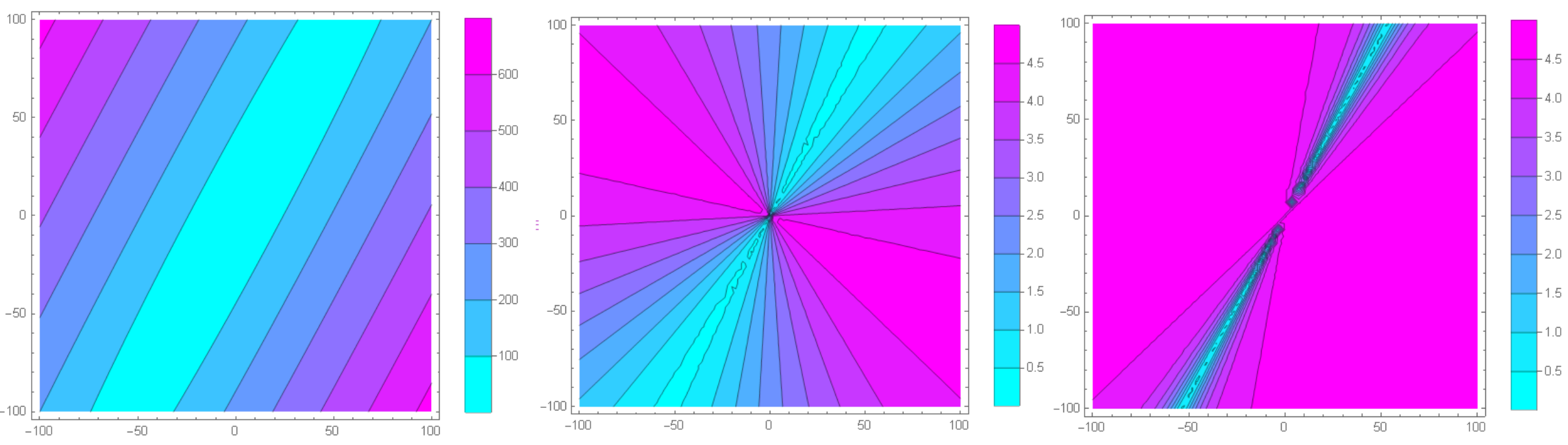} \\  
\begin{tabular}{ c c c }
 \includegraphics[width=0.162\textwidth]{ 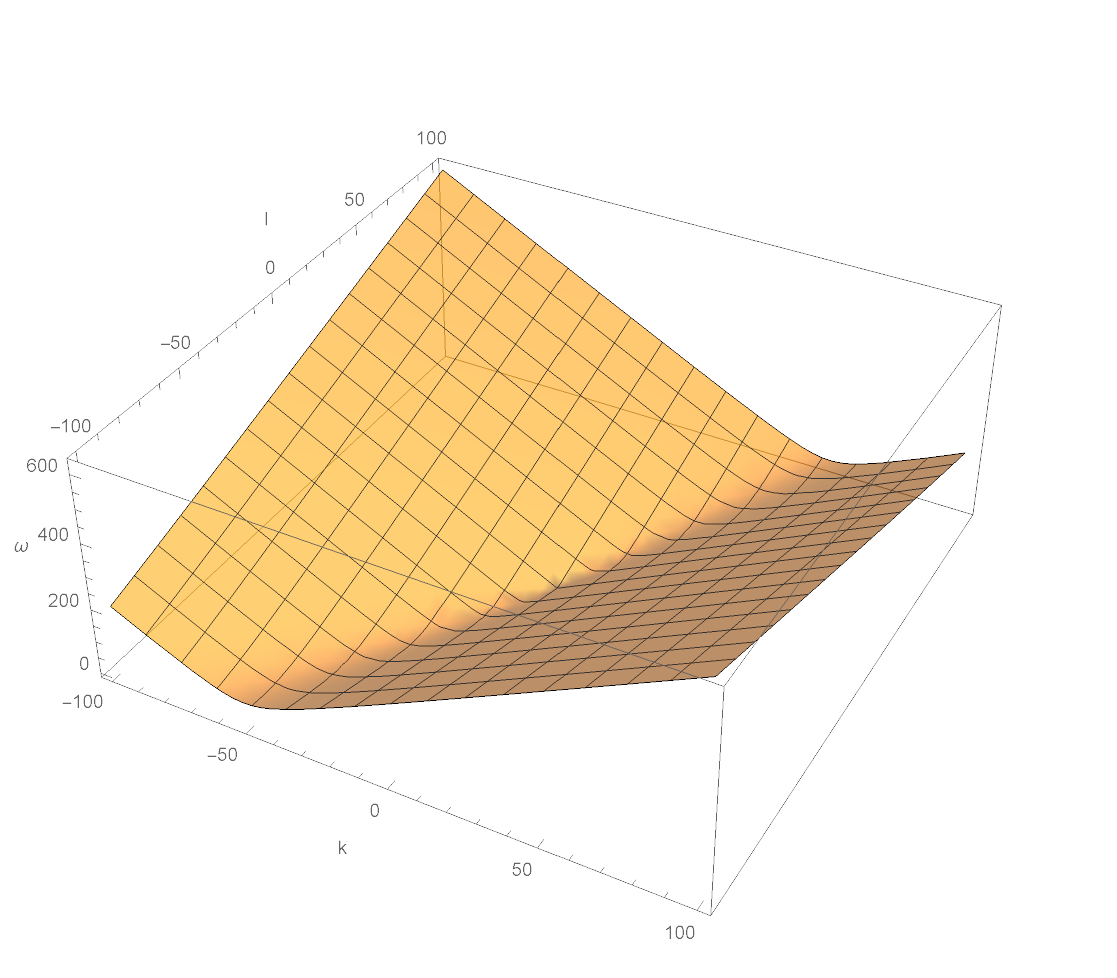} & \includegraphics[width=0.162\textwidth]{ 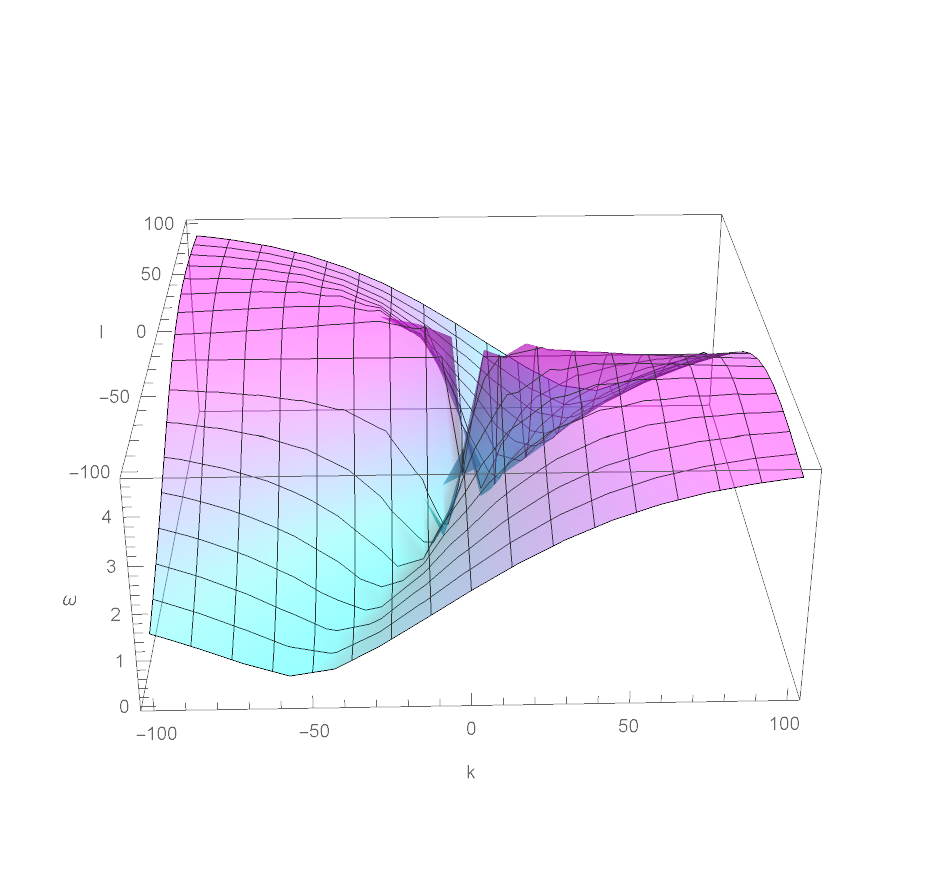} & \includegraphics[width=0.162\textwidth]{ 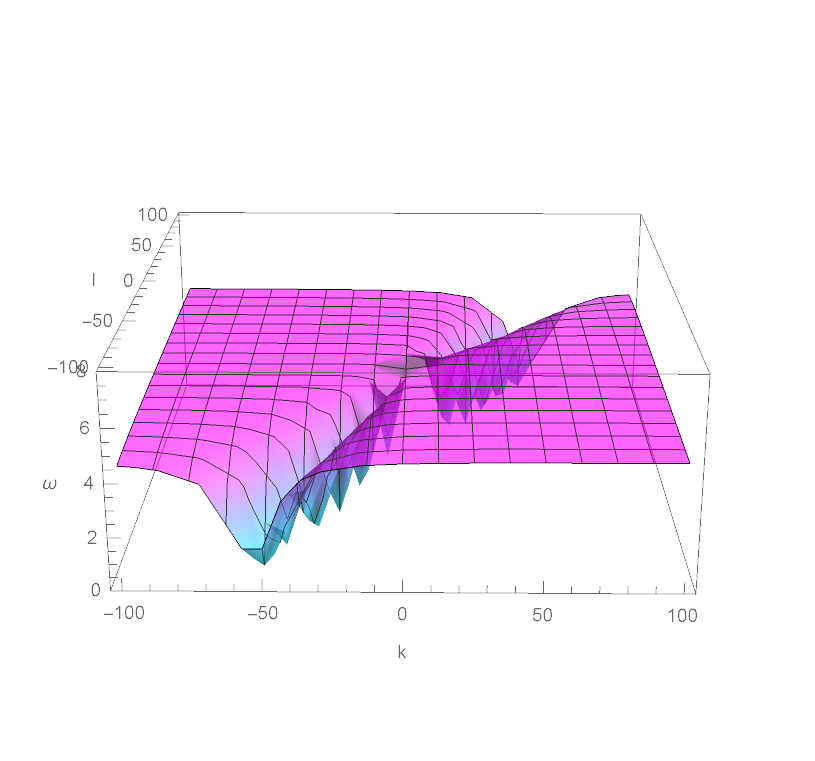}  
 \\
 $\omega$ & $v_f$ & $v_g$
\end{tabular}    
\end{tabular}
\end{center}
     \caption{In these graphics we represent the dispersion relation, phase speed and group speed  for a choice of parameters $R_9=1$, $R_{10}=2$, $m_1=n_1=n_2=1$, $m_2=2$ . In the top pictures we represent them in 2D and in the bottom the same quantities in 3D, since both complement visually the description. In the left column we observe dispersion relation $\omega(k,l)$ (Top contour plot and bottom as 3D plot), the central column represents those values for the phase speed. The right column corresponds to the group speed.}
    \label{fig:VelocidadadesContorno}
\end{figure}



When we impose the group velocity and the phase speed to be equal, generically the system exhibit dispersion. However, it also admits solutions without dispersion and non vanishing central charge. To illustrate it we give an example: for a choice of $j=n_r/m_r$ with a fixed $r$, and winding charges $m_r=m_s; n_r=-n_s$ with $r\ne s$ the two velocities coincide and the system does not exhibit dispersion. There are other more particular winding charge configurations that also satisfy the identity of velocities for the $j$ considered, like for example  $n_1=3, m_1=4, n=2=12, m_2=16$ with non-vanishing central charge. \newline

\paragraph{Second family of solutions} Since we are interested in allowing a more general dependence on time of the real scalar fields $A_r$, in order to obtain a generic field strength $\mathcal{F}$, we assume the following ansatz

{\small\begin{equation}
    Z_a= \left( \sin (\lambda(k \sigma +l \rho) )+  \cos (\lambda(k \sigma +l \rho) )\right)
    a_a e^{i \tau  \omega } \,,
\end{equation}
\begin{equation}
     A_{r}= a_{r} \left(a \sin (\lambda(k \sigma +l \rho) +\tau  \Omega )+b \cos (\lambda(k \sigma +l \rho) +\tau  \Omega )\right) \,.
\end{equation}}
See figure 1, that illustrates the $Z_a$ string QBL over the 2-torus. 
Proceeding as before, we obtain from the EOM for $Z$, the frequency $\omega$ is not affected by the modification of the $A_r$ ansatz. This is due to the fact that in the isotropic  case considered in this section, both EOM decouple and its value is,
\begin{equation}
\label{eq:RDispersiónZconArSintemporal}
  \omega^2=\lambda^2 \sum_r R_{r}^2 \left(k m_{r}-l n_{r}\right)^2 \,,
\end{equation}
and from the EOM for $A_r$, one obtains a different frequency $\Omega$. It depends on the moduli of the torus, the winding numbers and the amplitudes of the $A_r$ field, 
{\small
\begin{equation}
\label{eq:RDispersiónArconArSin}
\begin{aligned}
       \Omega^2 =& \lambda^2  \left(k m_{s}-l n_{s}\right)^2 R_{s}^2
    \\
    &- \lambda^2 
    \frac{a_{s}}{a_{r}}\left(k m_{r}- l n_{r}\right) \left(k m_{s}-l n_{s}\right) R_{r}R_{s} \, , 
\end{aligned}
\end{equation}
}
 with $r\ne s, r=1,2$.
In distinction with the previous case analyzed, the APD constraint is not automatically satisfied and leaves the following restriction,
\begin{equation}
  a_{r} R_{r} \left(k m_{r}-l n_{r}\right) +  a_{s} R_{s} \left(k m_{s}-l n_{s}\right) =0 \,.
\end{equation}
The constraint can be used to fix the relation $a=a_s/a_s$. Imposing it in (\ref{eq:RDispersiónArconArSin}) the frequency $\Omega$ becomes
\begin{equation}
    \Omega^2=2\lambda^2 R_{s}^2\left(k m_{s}-l n_{s}\right)^2  \,,
\end{equation}
for $s$ fixed. 
By imposing the constraint on  (\ref{eq:RDispersiónZconArSintemporal}),
\begin{equation}
    \omega^2=\lambda^2 R_{s}^2\left(k m_{s}-l n_{s}\right)^2 (1+a^2) \,,
\end{equation}
with no sum is intended. Due to the APD constraint and for this particular ansatz,  both frequencies become related by the amplitudes ratio of the $A_r$ field,
\begin{equation}
    \Omega^2=\frac{2\omega^2}{1+a^2} \,,
\end{equation}
This family of solutions allows for a nontrivial time dependence of $A_r$, resulting in a general symplectic curvature $\mathcal F$. Under the ansatz considered, the frequencies of the $A_r$ and $Z$ solutions become related, for $a=1$, both determine a unique excitation for a pair $(k,l)$. In this case, for a given moduli parametrizing the compact dimensions, there is a unique pair of Fourier modes for each set of winding numbers with a given central charge. The APD constraint represents the main restriction to the allowed solutions.
 
\paragraph{Dispersion} In order to  analyze the existence of solutions with vanishing dispersion relations, let us consider the case with $a=1$. It implies that the two components of the $A_r$ become equal and that $\Omega=\omega$. In distinction with the first family case, the relation between the Fourier modes of the non-compact sector $j=k/l$ gets restricted by the APD constraint to a unique value, fixed by the topological content, i.e. a fixed moduli and winding charges of the compact sector,
\begin{equation}\label{j}
j=k/l = \frac{(R_rn_{r}+R_{s}n_s)}{(R_{r}m_r+R_sm_s)} \,.
\end{equation}
The integer character of the Fourier modes forces, on general grounds,  the radii to be integer. 

Now, the velocity group associated to the dispersion relation is given by, 

\begin{equation}
\begin{aligned}
\textbf{v}_g=&\frac{\partial\omega}{\partial \textbf{k}} \equiv 
     \lambda R_{r} \sqrt{2}\left(m_{r}\ ,-    \ n_{r}\right);
    \quad\Rightarrow \\
     |\textbf{v}_g|= &\sqrt{2}\lambda R_{r} \sqrt{\left(m_{r}^2  +  n_{r}^2\right)} \ .   \end{aligned}
  \end{equation}
It has has a constant magnitude and the index  $r$ is fixed, there is no summation. 

Their phase speed is given by

\begin{equation}
\begin{aligned}
    c(\textbf{k})&= \sqrt{2}\lambda R_{r} 
    \frac{\left(k m_{r}- l n_{r}\right)}{\sqrt{k^2+l^2}} \, .
\end{aligned}
    \end{equation}
It is straightforward to see, that their phase speed along each of the directions verify  $jc_\sigma=c_\rho$.
We can observe that although the system presents generically dispersion, there exists areas which are dispersion-free. In order to see if there exists points in the phase space moduli free of dispersion, we impose $c(\textbf{k})=|\textbf{v}_g|,$
    \begin{equation}
 (k m_{r}-l n_{r})
     =
      \sqrt{k^2+l^2}\sqrt{\left(m_{r}^2  +  n_{r}^2\right)} \,,
\end{equation}
which implies,
\begin{equation}\label{condicion_no_dispersion}
  l=-\frac{n_{r}}{m_{r}} k \ . 
\end{equation}
 for $r$ fixed. Hence, for central charge different from zero the system in general exhibits dispersion but it also allows solutions with no dispersion satisfying (\ref{condicion_no_dispersion}).

A particular subfamily of solutions corresponds to those that appear by imposing $\partial^2 f\propto f$. This solution subfamily corresponds to the  $\lambda=1$ choice.
The main constraints on this first family of solutions are imposed by the periodicity condition required since they propagate on the world volume of a compact membrane. Solving this system of equations on a flat plane would admit much more general solutions than those considered here, for the same given ansatz.\newline

 We have found two  string-like families with a non-vanishing cost in energy solutions that satisfy the Q-ball-ansatz to the nonlinear system. The non-linearity is present in spite of the vanishing symplectic bracket due to the ansatz choice of the symplectic field $\mathbb A$ and to the complex scalar field $Z$ which become determined in terms of a same real single function $f(\sigma, \rho)$. The system is highly simplified  but still non-linear, and it admits solutions sinusoidal type.  It is important to mention that these solutions are not present for the case without worldvolume fluxes, they exist due to them. They correspond to excitations with minimal energy. In addition, as we can observe when represented over the 2-torus, they generate patterns associated with twists over the surface. They admit solutions with and without dispersion. \newline

\paragraph{Localized dynamic QBL solutions over the worldvolume} Now, we would like to show that the M2-brane with worldvolume fluxes also admits a family of string-like solutions that correspond to localized spinning Q-ball-like solutions propagating over the worldvolume that rotate and have the characteristic Q-ball breathing modes. By localization we mean a localized pulse over the torus worldvolume defined by $f(\sigma,\rho,\tau)$ for  a fixed time $\tau_0$. We illustrate a particular example but there can be found many more. The solutions evolve in time as shown in figure 3. It also exhibits in the phase the characteristic dependence of a spinning Q-ball. This solution may be also seen as the Lorentz of a boost of a static QBL solution.
\newline
In order to find these solutions we generalize the ansatz to the following one:
\begin{equation}
 Z_a=f_a(\sigma,\rho,\tau)e^{i(\tau  \omega+k \sigma +l \rho)}
 \end{equation}
  with $f_a$ being real, acting in the general equations given by (
\ref{equationofmotionnoncompactcomplexZ},\ref{equationofmotionAcomplex},\ref{constraint_complex_general}) and subject to the topological constraint (\ref{central charge})  
.
An admissible string-like family of localized solutions of solitonic profile is the following  one,
\begin{equation}
    Z_a=f_a(\sigma,\rho,\tau)e^{i\beta}= r_a Arcotan( a e^{ b \sin (\beta )+ b \cos (\beta )})  e^{i\beta} \,,
\end{equation}
\begin{equation}
     A_{r}= a_{r} \left( \sin (k \sigma +l \rho +\tau  \Omega )+ \cos (k \sigma +l \rho +\tau  \Omega )\right) \,.
\end{equation}
where 
\begin{equation}
   \beta=k\sigma+l\rho+\omega\tau
\end{equation}
This string-like solution correspond to a dynamic rotating Q-ball with non-vanishing breathing modes, that propagates along the worldvolume. For a visual representation, see Figures \ref{fig:LocalRho} and \ref{fig:LocalToro}. In order to illustrate better the excitation, only the dependence of the pulse on one of the variables is shown.
\begin{figure}[htbp]
    \centering
    \begin{tabular}{c}
\animategraphics[loop,autoplay,width=0.25\textwidth]{5}{giffLocalSolution-Rho2D/LocalRho2d}{0}{99} 
\\  
\animategraphics[loop,autoplay,width=0.25\textwidth]{5}{giffLocalSolution-Rho2D/LocalRho3d-}{0}{99} 
\\  
\animategraphics[loop,autoplay,width=0.25\textwidth]{12.5}{giffLocalSolution-Rho2D/LocalRhoToro-}{0}{249} 
    \end{tabular}
    \caption{
    \label{fig:LocalRho}
    Using the following coefficients
    $ n_1  = 1 \ n_2 \ = 0 \, m_1  = 0 \, m_2  = 1 \, R_1  = 1 \, R_2  = 2 \, n  = 1, \ k  = 0 \, l  = 1 \, \omega  = 1 \, \phi  = -(1/4)\pi , \ r_a=1, a=0.0001 \ and\  b=20$
    }
\end{figure}

\begin{figure}[htbp]
    \centering
    \begin{tabular}{c}
\animategraphics[loop,autoplay,width=0.25\textwidth]{5}{GiffLocalSolution/Local2d}{0}{99}
\\  \animategraphics[loop,autoplay,width=0.25\textwidth]{12.5}{GiffLocalSolution/LocalToro}{0}{249}
    \end{tabular}
    \caption{
    \label{fig:LocalToro}
    The Cartesian coordinates solution is displayed at the top, with the solution plotted on the torus shown at the bottom. Using the following coefficients
    $ n_1  = 1 \ n_2 \ = 0 \, m_1  = 0 \, m_2  = 1 \, R_1  = 1 \, R_2  = 2 \, n  = 1 \ k  = 1 \, l  = 1 \, \omega  = \sqrt{5} \, \phi  = -(1/4)\pi , \ r_a=1, a=0.0001\ and\ b=20$}
\end{figure}
It's worth noting that the frequency remains unchanged compared to the previously discussed solutions. This is due to the decoupling of equations of motion (EOM) in the isotropic cases considered in this section. By following the same analysis done for the second family of solutions the frequencies $\omega$ and $\Omega$ can be obtained and they correspond to the same values previously obtained, i.e. for $a=1$ and fixed $s$.
\begin{equation}
    \Omega^2=2 R_{s}^2\left(k m_{s}-l n_{s}\right)^2 =\omega^2 \,,
\end{equation}
 
The dispersion relation is the same as the one previously analyzed, hence also admits dispersion-free solutions by imposing $n_s$ multiple of $m_s$ for a fixed $s$.


\section{Numerical Membrane Q-ball-like solutions}  
In this section we consider the non-isotropic case with the nonlinearities associated with the complex scalars $Z_a$ present, and not just induced because of the coupling to the $A_r$ scalar. They still satisfy the ansatz $\partial_\sigma f(\sigma, \rho)\propto \partial_\rho f(\sigma,\rho)$. These solutions are numerical. For simplicity, we will assume a vanishing symplectic field $\mathbb{A}$, i.e. $A_r$ is assumed to be constant or with a linear dependence only on $\tau$, focusing exclusively in the role of the scalars $Z_a$.

\subsection{Q-ball-like ansatz for the isotropic case} 
As a warm-up exercise to test the numerical computation error,  in the isotropic case, we compare it with previously obtained analytical results.  
 For the isotropic case and constant $A_r$ we consider the system of equations given by (\ref{System1}) and (\ref{System2}) subject to the central charge condition \ref{cargaCentral} discussed in Section 4. 
The Q-ball ansatz automatically satisfy the Diffeomorphism area preserving constraint.
The solutions of the equation (\ref{System1}) was analyzed in \cite{SpinningSolutionsBosonicM2-brane2022} and a discrete set of frequencies was obtained.  From a  numerical point of view the M2-brane system of partial differential equations that we consider is:  i) {\it Overdetermined}. In the sense that the unknown function must satisfy two different differential equations ii) {\it Nonlinear}. Although only to the second equation, where in this case the non-linearities are represented by products of different derivatives (in order and variable) of the unknown function and iii){\it Rare}. The most striking feature is that this form of nonlinearity is unusual and, as far to our knowledge, there are no reported numerical methods for solving it. Neither in the finite dimensional (ODEs) or infinite dimensional (PDEs) case.
The three steps algorithm we will follow in order to a find numerical solution to the (\ref{System1}-\ref{System2}
) system will be

\begin{itemize}
	\item[i.] Set the boundary condition.
	In principle boundary ($\Gamma$) condition can be fixed as: Dirichlet or periodic, though in this case we will restrict ourselves to periodic boundary conditions on $\Gamma$:
	\begin{eqnarray}
	f(0,\rho) = f(2 \pi, \rho), \quad 
	f(\sigma,0) = f(\sigma,2 \pi) \,.
	\end{eqnarray}
	\item[ii.] Solve (\ref{System1} 
 ) as an eigenvalue problem:
	\begin{equation}
	{\cal L}_{p} \varphi_n(\sigma,\rho) = \omega_n ~ \varphi_n(\sigma,\rho).
	\end{equation}
	\item[iii.] Write the solution of (\ref{System2}
 ) as
	\begin{equation}
	f_{a}(\sigma,\rho) = \sum_{n=0}^{\infty} \gamma_n\varphi_n(\sigma,\rho).
	\label{Solution}
	\end{equation}
	and solve an optimization problem to find the best set of $\gamma_n$, so that (\ref{Solution}) is solution of (\ref{System2}
 ).  
\end{itemize}
\subsubsection{Finite differences approach for the solution} The system (\ref{System1})-(\ref{System2}) can be re-written using centered Euler scheme: 

\begin{eqnarray}
{\cal D}_{\sigma} \varphi_n &=& \frac{\varphi_n^{i+1,j}- \varphi_n^{i-1,j}}{2 \Delta \sigma},
\\
{\cal D}_{\rho} \varphi_n &=& \frac{\varphi_n^{i,j+1}- \varphi_n^{i,j-1}}{2 \Delta \rho},
\\
{\cal D}^2_{\sigma} \varphi_n &=& \frac{\varphi_n^{i+1,j}-2 \varphi_n^{i,j} + \varphi_n^{i-1,j}}{\Delta \sigma^2}
\\
{\cal D}^2_{\rho} \varphi_n &=& \frac{\varphi_n^{i,j+1}-2 \varphi_n^{i,j} + \varphi_n^{i,j-1}}{\Delta \rho^2},
\end{eqnarray}
\noindent
with the mixed second derivative given:

\begin{equation}
{\cal D}^2_{\sigma,\rho} \varphi_n = \frac{\varphi_n^{i+1,j+1} - \varphi_n^{i-1,j+1} - \varphi_n^{i+1,j-1} + \varphi_n^{i-1,j-1}}{ 4 \Delta \sigma \Delta \rho}
\end{equation}

\noindent
It is easy to verify that:

\begin{equation}
\begin{aligned}
& {\cal D}_{\rho} \approx {\partial}_{\rho}, \quad {\cal D}_{\sigma}  \approx  {\partial}_{\sigma},\quad  {\cal D}^2_{\sigma} \approx {\partial}^2_{\sigma}, \\
 &{\cal D}^2_{\sigma,\rho} \approx {\partial}^2_{\sigma,\rho}, 
\quad  {\cal D}^2_{\rho} \approx  {\partial}^2_{\rho} \,. 
\end{aligned}  
\end{equation}

Since equation (\ref{System1}) is linear, it can be solved relatively easily using finite differences or built-in functions from Mathematica or Python. In the case of the second equation we must use a different methodology due to its intrinsic nonlinearities.
\subsubsection{A finite differences representation for the nonlinear equation}
Re-writing the equation (\ref{System2}) in terms of finite differences become,
{\fontsize{6}{6}
\begin{equation}
\begin{aligned}
&\left(m_{r} \partial_{\sigma}^{2} \sum_{n=0}^{\infty} \gamma_n\varphi_n(\sigma,\rho)-n_{r} \partial^2_{\sigma \rho} \sum_{n=0}^{\infty} \gamma_n\varphi_n(\sigma,\rho)\right) \partial_\rho \sum_{n=0}^{\infty} \gamma_n\varphi_n(\sigma,\rho)+\\
&-\left(m_{r} \partial^2_{\rho \sigma} \sum_{n=0}^{\infty} \gamma_n\varphi_n(\sigma,\rho)-n_{r} \partial_{\rho}^{2} \sum_{n=0}^{\infty} \gamma_n\varphi_n(\sigma,\rho) \right)\partial_{\sigma} \sum_{n=0}^{\infty} \gamma_n\varphi_n(\sigma,\rho) = 0. \\ 
\end{aligned}
\end{equation}

\begin{eqnarray}
\left(m_{r}  \sum_{n=0}^{\infty} \gamma_n \partial_{\sigma}^{2} \varphi_n(\sigma,\rho)-n_{r}  \sum_{n=0}^{\infty} \gamma_n \partial^2_{\sigma \rho} \varphi_n (\sigma,\rho)\right)  \sum_{n=0}^{\infty} \gamma_n \partial_\rho \varphi_n(\sigma,\rho) &-& \nonumber \\
\left(m_{r}  \sum_{n=0}^{\infty} \gamma_n \partial^2_{\rho \sigma} \varphi_n(\sigma,\rho)-n_{r} \sum_{n=0}^{\infty} \gamma_n \partial_{\rho}^{2} \varphi_n(\sigma,\rho) \right) \sum_{n=0}^{\infty} \gamma_n \partial_{\sigma} \varphi_n(\sigma,\rho)  &=& 0. \nonumber \\
\end{eqnarray}

\begin{eqnarray}
\left[ \sum_{n=0}^{\infty} \gamma_n \left( m_{r} \partial_{\sigma}^{2} \varphi_n(\sigma,\rho)- n_{r} \partial^2_{\sigma \rho} \varphi_n (\sigma,\rho) \right) \right] . \sum_{n=0}^{\infty} \gamma_n \partial_\rho \varphi_n(\sigma,\rho) &-& \nonumber \\
\left[\sum_{n=0}^{\infty} \gamma_n \left( m_{r} \partial^2_{\rho \sigma} \varphi_n(\sigma,\rho)-  n_{r} \partial_{\rho}^{2} \varphi_n(\sigma,\rho) \right) \right] . \sum_{n=0}^{\infty} \gamma_n \partial_{\sigma} \varphi_n(\sigma,\rho)  &=& 0. \nonumber \\
\end{eqnarray}
}

\vspace{0.5cm}

This can be re-expressed in a more compact form as:

\begin{eqnarray}
\sum_{n=0}^{\infty} \gamma_n \left( A_n - B_n \right) \cdot \sum_{n=0}^{\infty} \omega_n C_n + \nonumber \\-
\sum_{n=0}^{\infty} \gamma_n  \left(B_n - D_n \right) \cdot \sum_{n=0}^{\infty} \omega_n E_n  &=& 0.
\label{matrix-version}
\end{eqnarray}
%
%
%
where: 
\begin{eqnarray}
A_n &=& m_{r} \partial_{\sigma}^{2} \varphi_n(\sigma,\rho), \nonumber \\
B_n &=& n_{r} \partial^2_{\sigma \rho} \varphi_n (\sigma,\rho), \\ 
C_n &=& \partial_\rho \varphi_n(\sigma,\rho) \nonumber    
\end{eqnarray}

and
\begin{equation}
D_n = \partial_{\rho}^{2} \varphi_n(\sigma,\rho), \quad 
E_n = \partial_{\sigma} \varphi_n(\sigma,\rho) \,.
\end{equation}

Using the Cauchy product in (\ref{matrix-version}), we obtain:

\begin{eqnarray}
\sum_{n=0}^{\infty} \sum_{m=0}^{n} \gamma_n \left( A_n - B_n \right) ~ \odot\gamma_{n-m} C_{n-m} - \nonumber \\
\sum_{n=0}^{\infty} \sum_{m=0}^{n} \gamma_n  \left(B_n - D_n \right) ~ \odot\gamma_{n-m} E_{n-m}  = 0 \,.    
\end{eqnarray}

\noindent

\begin{equation}
\sum_{n=0}^{\infty} ~\gamma_n \left( \sum_{m=0}^{n} \gamma_{n-m} S_{n,m} \right)  = 0,
\label{rare-equation}
\end{equation}

\noindent
with, 
\begin{equation}
S_{n,m} = \left( A_n - B_n \right) \odot  C_{n-m} -  \left(B_n - D_n \right) \odot E_{n-m} \,, 
\end{equation}
 and $\odot$ represents the pointwise product. 
Thus, the solution of the problem can be reduced to find a solution for this nonlinear algebraic equation with matrix coefficients and $\gamma_n$ as unknowns.
\onecolumngrid
\begin{center}
    \begin{figure}
	\centering
\includegraphics[width=0.25\textwidth]
  { 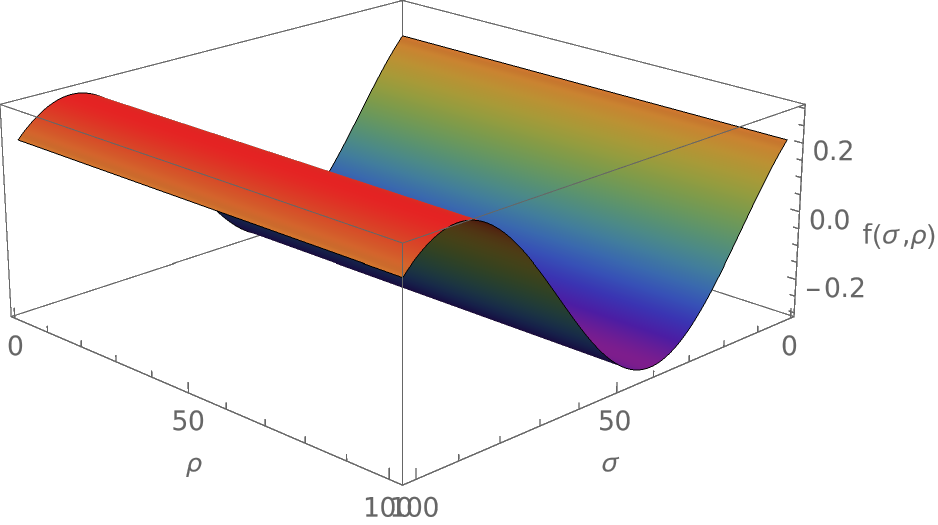}~~
		\includegraphics[width=0.25\textwidth]
  { 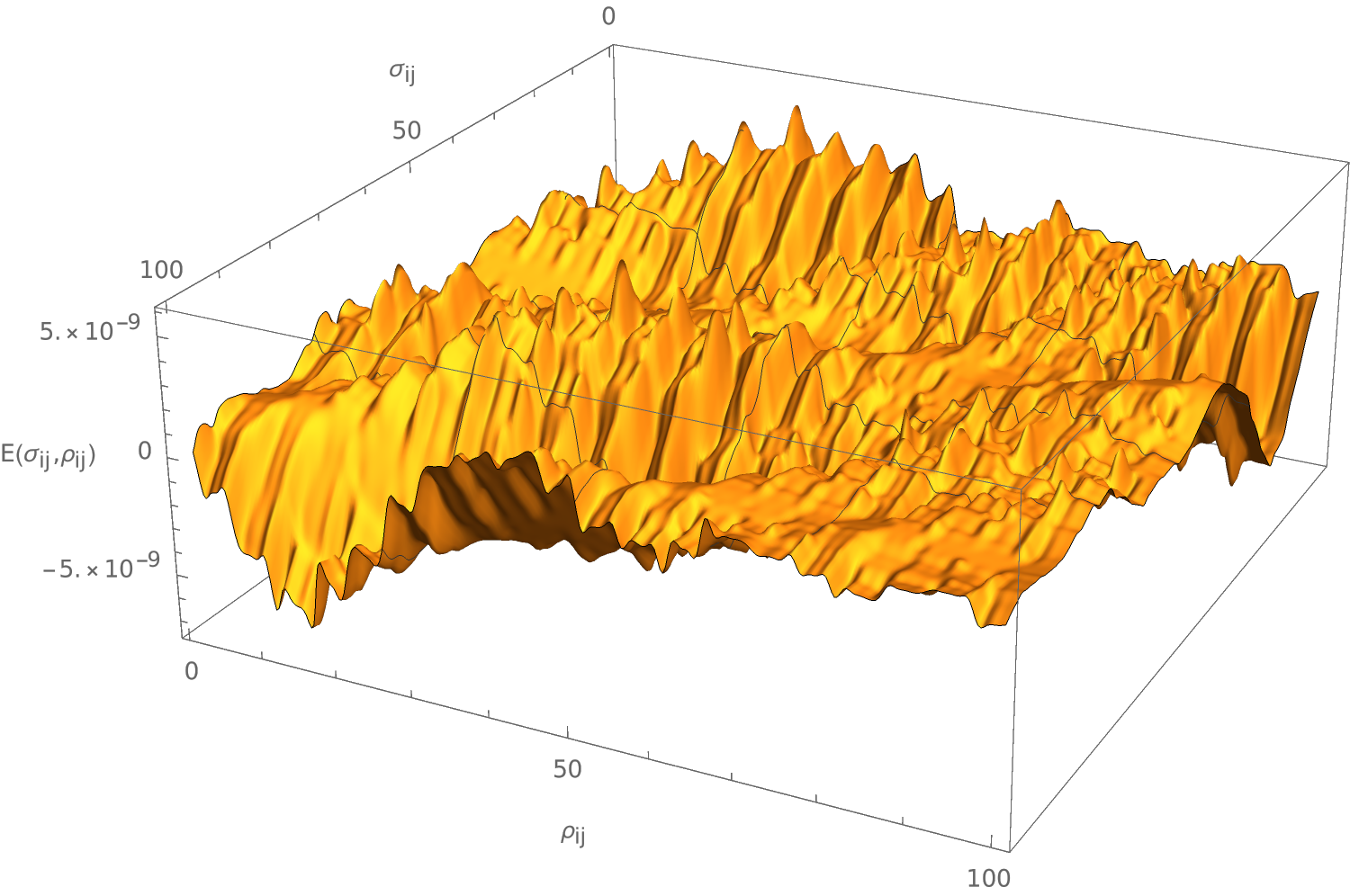}\\
		\includegraphics[width=0.25\textwidth]
  { 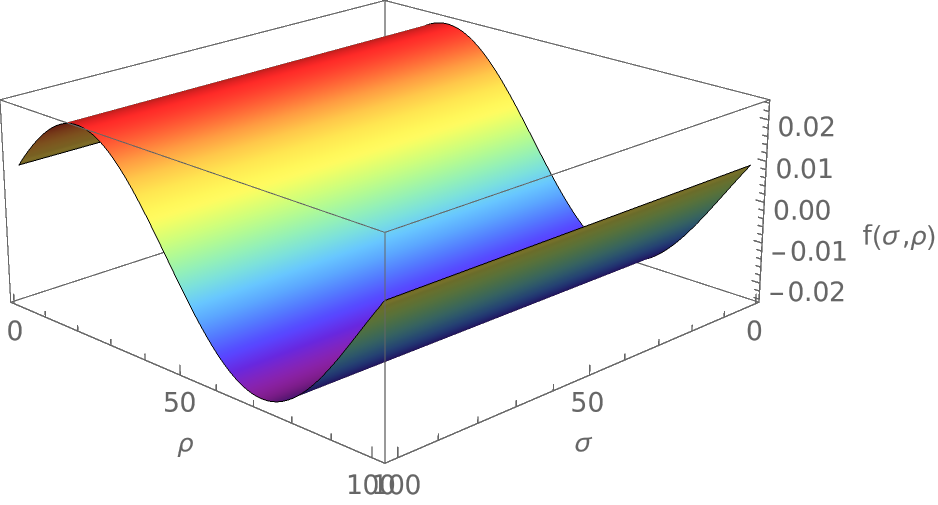}~~
		\includegraphics[width=0.25\textwidth]
  { 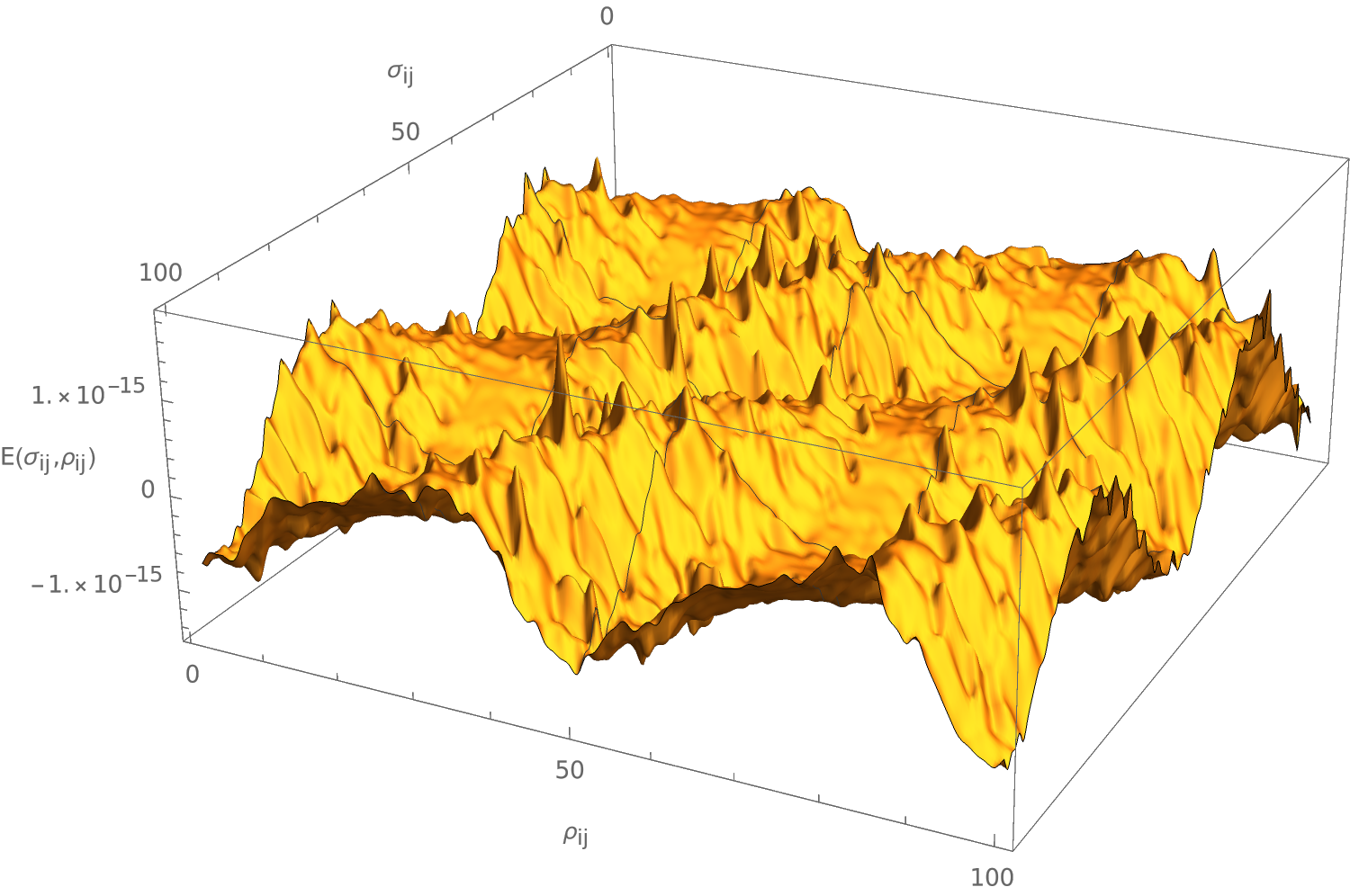}\\
		\includegraphics[width=0.25\textwidth]
  { 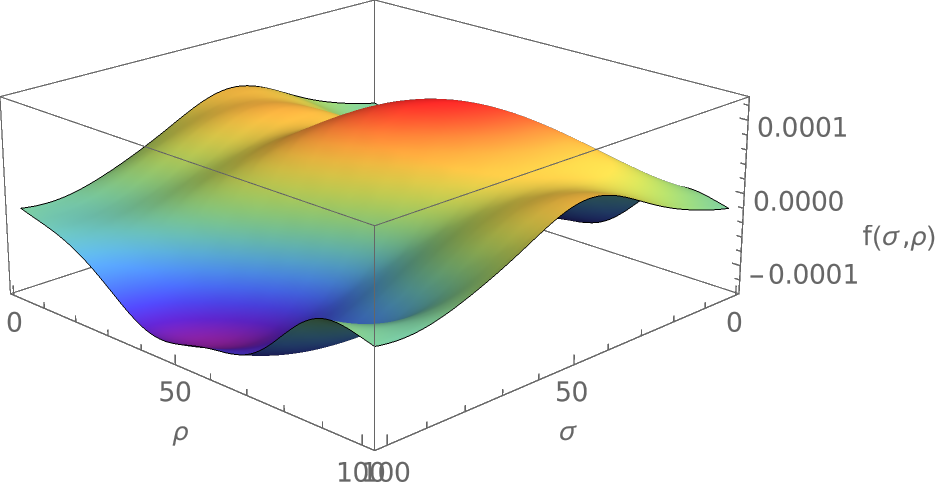}~~
		\includegraphics[width=0.25\textwidth]
  { 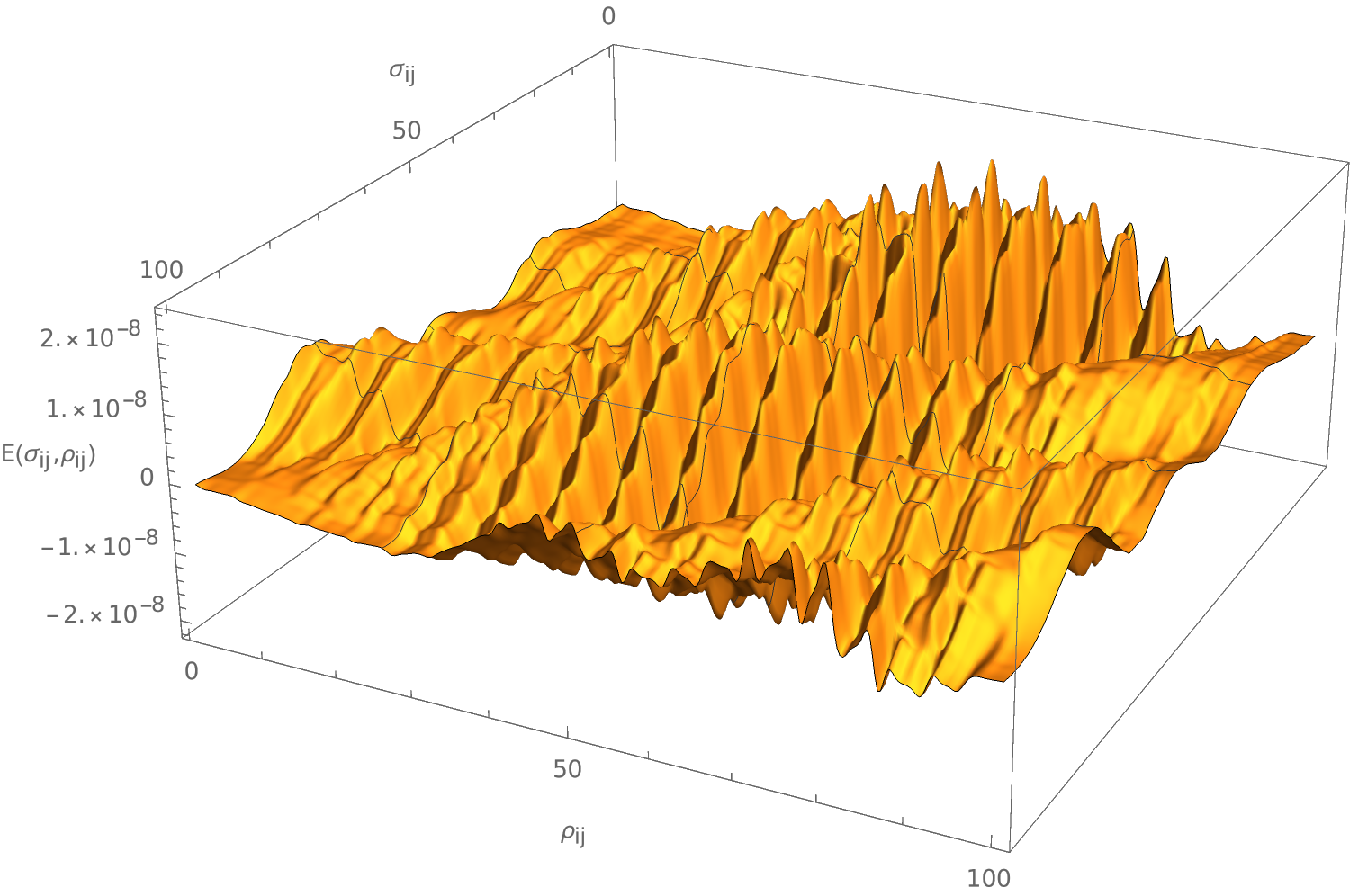}\\
		\includegraphics[width=0.25\textwidth]
  { 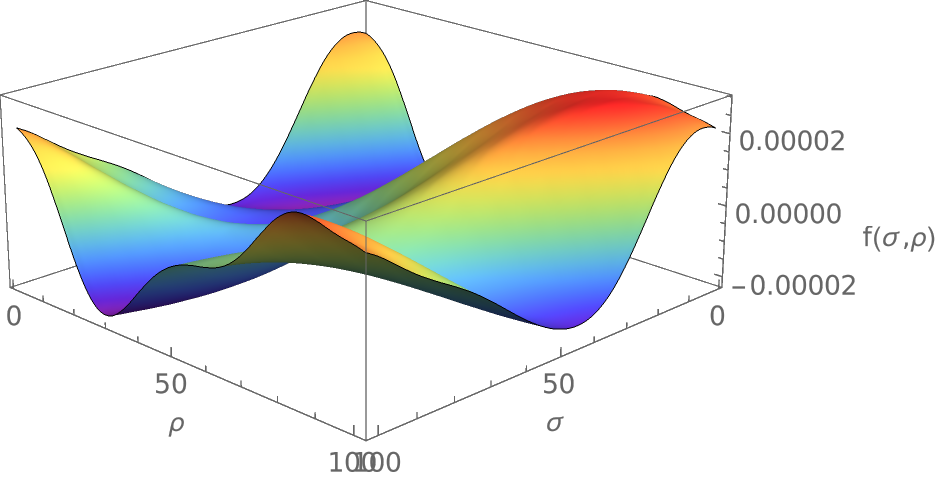}~~
		\includegraphics[width=0.25\textwidth]
  { 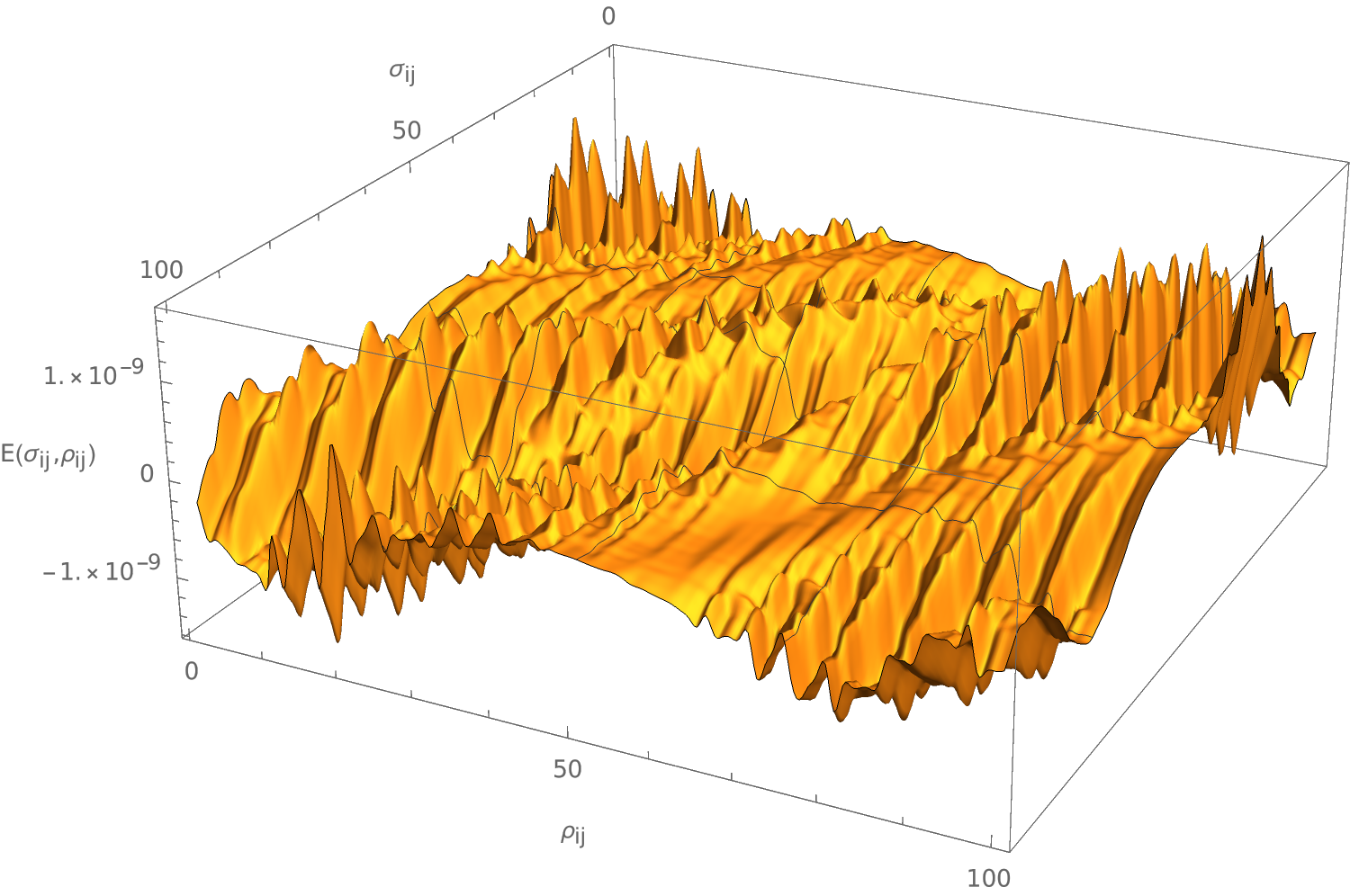}
	\caption{\label{approximateSolution} Approximate solution of (\ref{System1})-(\ref{System2}) and approximation error.}
\end{figure}
\end{center}

\twocolumngrid

We define the error function in the following way,
{\small
\begin{equation}
\begin{aligned}
E= \left( m_{r} \partial_{\sigma}^{2} f_{a}(\rho_{ij}, \sigma_{ij})-n_{r} \partial^2_{\sigma \rho} f_{a}(\rho_{ij}, \sigma_{ij}) \right) \partial_\rho f_{a}(\rho_{ij}, \sigma_{ij}) - \\
 \left(m_{r} \partial^2_{\rho \sigma} f_{a}(\rho_{ij}, \sigma_{ij})-n_{r} \partial_{\rho}^{2} f_{a}(\rho_{ij}, \sigma_{ij}) \right) \partial_{\sigma} f_{a}(\rho_{ij}, \sigma_{ij}) \,,
\end{aligned}
\end{equation}
}

\noindent with $E\equiv E(\rho_{ij},\sigma_{ij})$. 
Figura \ref{approximateSolution} shows graphs of the possible solutions (note that the numerical scheme allows for many, unless the number of bases equals the number of lattices squared) and the errors made in those approximations at each lattice point.

We obtain two kinds of solutions: some with large amplitudes and others with small amplitudes.  This may be because the parameters $n$, $m$, and $R$ may be in very different ranges. 


\subsection{Membrane Q-ball-like solution: The non-isotropic case }
In section 3. we have obtained the first exact analytical families of solutions for a Q-ball ansatz of the complex scalar field. They correspond to isotropic solutions  and with the nonlinear behaviour encoded in the $A_r$ field which is expressed in terms of $f$, and hence makes the complex scalar field $Z$ to be nontrivial. However, the family of solutions obtained trivializes the non-isotropic contribution of the brackets to the $Z$ EOM. In order to capture their contribution, in this section we consider the non-isotropic case in the absence of a symplectic gauge field $\mathbb{A}$. 

Imposing  $A_{r} = Constant$ in the E.O.M (\ref{EOMZQballfamilia}) they become the following nonlinear but simplified system of equations:
\begin{eqnarray}
\label{eq:fNoIsotropico}
\omega_c^2 f_c=-
\sum_{r=9}^{10}
\left[
R_{r}(j_c m_r -n_r)
\right]^{2}
\partial_{\rho}^{2}f_c + \nonumber\\
-\sum_{a} (j_c-j_a)^2             
\partial_{\rho}^{2} f_{c}
(\partial_\rho f_a )^2 \  .
\end{eqnarray}
with $a,c=1,2,3$ and $r,s=1,2$. In order to simplify the notation, let us define the constants $\kappa_a$,
\begin{equation}
\kappa_{a} = \sum_{r=9}^{10}
\left[
R_r(j_a m_r -n_r)
\right]^{2} \,,
\end{equation}
\noindent
in addition to the matrices $\lambda$ and $\mu$, with elements:
\begin{equation}
\lambda_{c,a} = (j_c - j_a)^2, \qquad
\mu_{r,a} = \left[
R_{r} \left( j_a m_{r}-n_{r}\right) \right]^2 \,,
\end{equation}
%
with $r$, $c$ y $a$, defined as before. With this notation, the above equations can be written as:

\begin{eqnarray}
- \kappa_{1}~
\partial_{\rho}^{2}f_1 
-
\partial_{\rho}^{2}f_1  ~ \sum_{a} \lambda_{1,a}             
(\partial_\rho f_a )^2 &=&  \omega_1^2 f_1  \,, \nonumber \\
- \kappa_{2}~
\partial_{\rho}^{2}f_2 
-           
\partial_{\rho}^{2}f_2  ~ \sum_{a} \lambda_{2,a}    
(\partial_\rho f_a )^2 &=& \omega_2^2 f_2  \nonumber \,,   \\
- \kappa_{3}~
\partial_{\rho}^{2}f_3 
-
\partial_{\rho}^{2}f_3  ~ \sum_{a} \lambda_{3,a}             (\partial_\rho f_a )^2 &=& \omega_3^2 f_3   \,.
\label{system}
\end{eqnarray}
%
%

\subsubsection{A numerical approach to the solution} 

In order to write our problem in a convenient way for its numerical integration, $x' = {\mathbf f}(x)$ con  $x \in \Re^3$, let's start by defining $\partial_{\rho} f_a = g_a$, so that the equations (\ref{eq:fNoIsotropico}) can be written as:
\begin{eqnarray}
&\partial_{\rho} f_1 = g_1 ; \quad\nonumber 
- \kappa_{11}~
\partial_{\rho} g_1 
-
\partial_{\rho} g_1  ~ \sum_{a} \lambda_{1,a}             
g_a^2 = \omega_1^2 f_1  \,, \nonumber \\
&\partial_{\rho} f_2 = g_2; \quad\nonumber 
- \kappa_{22}~
\partial_{\rho}  g_2 
-           
\partial_{\rho}  g_2  ~ \sum_{a} \lambda_{2,a}             
g_a^2 = \omega_2^2 f_2 \,, \\
&\partial_{\rho} f_3 = g_3; \quad 
- \kappa_{33}~
\partial_{\rho} g_3 
-
\partial_{\rho} g_3  ~ \sum_{a} \lambda_{3,a}             
g_a^2 = \omega_3^2 f_3  \,.
\label{system-ODE}
\end{eqnarray}
%
%
%
So, if we include the boundary conditions, our problem will look like:
\begin{eqnarray}
\partial_{\rho} f_b = g_b;  \qquad
\partial_{\rho} g_b  =  \frac{\omega_b^2 f_b}{\kappa_{bb}~+~ \sum_{a} \lambda_{b,a} g_a^2} \,,
\end{eqnarray}
%

\noindent
with $f_i(0) = f_i(2 \pi) \quad i=1,2,3$. Thus, our problem is to find a function:
\begin{equation}
x(\rho) = ( f_1(\rho), g_1(\rho), f_2(\rho), g_2(\rho), f_3(\rho), g_3(\rho) ),
\end{equation}
%
that satisfies:
\begin{equation}
\frac{\partial  x(\rho)}{\partial \rho} = {\mathbf f}(x) \,,
\end{equation}
 %
where ${\mathbf f}$ is a vector-valued function given by:
\begin{equation}
{\mathbf f}(x) =
\left(
\begin{array}{c}
 g_1(\rho) \\ \\
 \frac{\omega_1^2 f_1(\rho)}{\kappa_{1}~+~ \sum_{a} \lambda_{1,a} g_a^2(\rho)} \\\\
   g_2(\rho) \\\\
  \frac{\omega_2^2 f_2(\rho)}{\kappa_{2}~+~ \sum_{a} \lambda_{2,a} g_a^2(\rho)} \\\\
g_3(\rho)\\\\
 \frac{\omega_3^2 f_3(\rho)}{\kappa_{3}~+~ \sum_{a} \lambda_{3,a} g_a^2(\rho)}
\end{array}
\right),
\label{Dynamical-System}
\end{equation}
%
with the boundary condition $x(0)= x(2 \pi)\label{BC}$.

\subsubsection{The proposed solution} 
Since this problem can be assimilated to the class of "two point boundary value problems", we propose to use a customized version of the "shooting method", to obtain numerical solutions.
In these methods, since only part of the initial condition ($f_i(0)$) is known, the remaining part is assumed to be known and the initial value problem is solved, varying this last part, until the solution complies with the remaining boundary condition.

A more or less efficient way to implement this idea, is to find the roots ($g_1^*,g_2^*,g_3^*$) of the function:
\begin{equation}
F(x(0),\rho) = \Phi(x(0),2 \pi) - x(2 \pi) \,,
\label{problem}
\end{equation}
%
where $\Phi(x(0),\rho)$ is the solution of the initial value problem:
\begin{equation}
\frac{\partial  x(\rho)}{\partial \rho} = {\mathbf f}(x) \,,
\end{equation}
%
with initial conditions $$x(0) =(f_1(0),g_1(0),f_2(0),g_2(0),f_3(0),g_3(0))$$.

The most popular way to solve the problem (\ref{problem}), is using Newton's method. This type of methods can be extended to the case of problems of the type that concern us ("two point boundary values") in many ways. One of them, given in \cite{Aprille}, states that:
\begin{equation}
x^{i+1}(0) = x^{i}(0) + (J_{\Phi} -I)^{-1} \left[\Phi(x^{i}(0),2 \pi)- x^{i}(0) \right] \,,
\end{equation}
%
where $J_{\Phi}$ is the derivative of the solution ($\Phi$) evaluated $2 \pi$, also called "sensitivity matrix",
\begin{equation}
J_{\Phi} = \frac{\partial \Phi(x(0),2 \pi)}{\partial x(0)}|_{x^i(0)} \,.
\label{jacobian}
\end{equation}

\noindent
and $I$ is the identity matrix.

In this case (\ref{jacobian}) it is estimated using the simplest of the strategies, i. e., the problem system is solved in the case of two close initial conditions $x^i(0)$ and $x^{i}(0)+h$ $\Phi(x^{i}(0 ),2 \pi)$ and $\Phi(x^{i}(0) + \delta, 2\pi)$ and $J_{\Phi}$ is approximated as:

\begin{equation}
J_{\Phi} \approx \frac{\Phi(x^{i}(0),2 \pi)-\Phi(x^{i}(0) + \delta, 2\pi)}{\delta}.
\end{equation}

On the other hand, the solution $\Phi(x^{i}(0),2 \pi)$ is estimated using Euler's method, that is, a discretization (the simplest) of the form:
\[
x_{n+1} = x_{n} + h \, \mathbf{f}(x_{n})
\]

\begin{figure}[ht]
	\begin{center}
\includegraphics[width=0.45\textwidth]
  { 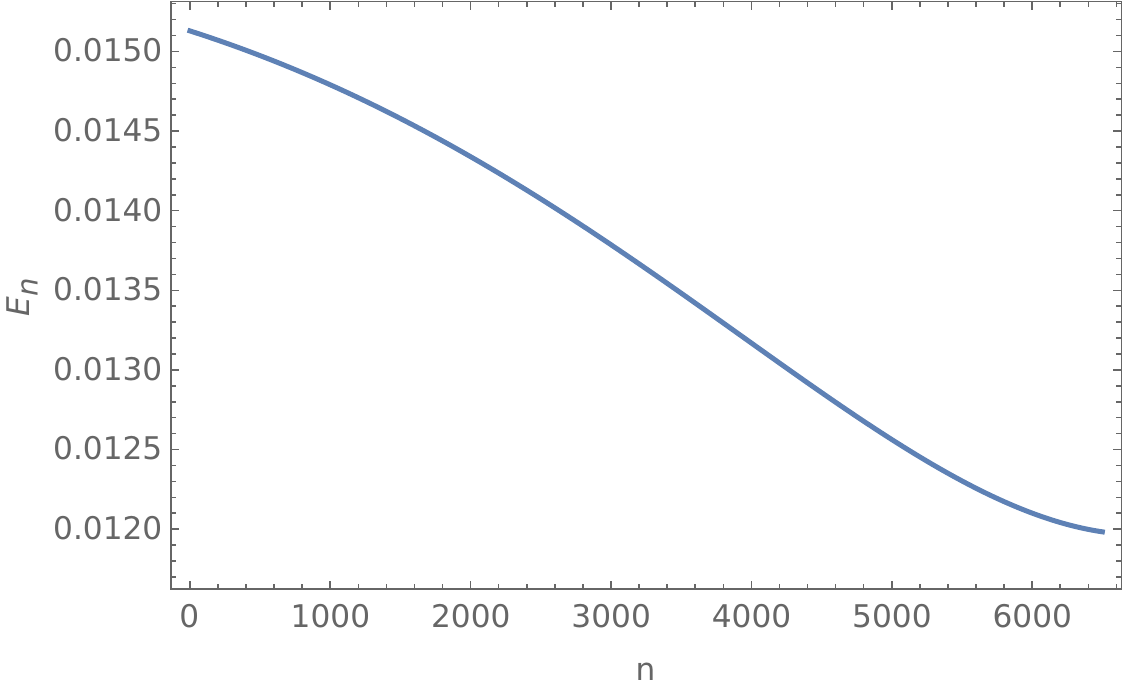}
	\end{center}
	\caption{	\label{Fig1} $E_n$ represents the difference $x(0) - \Phi(x(0),2\pi)$, in each iteration of Newton's method.}
\end{figure}

\begin{figure}[ht]
	\begin{center}
		\includegraphics[width=7cm, height=6cm]{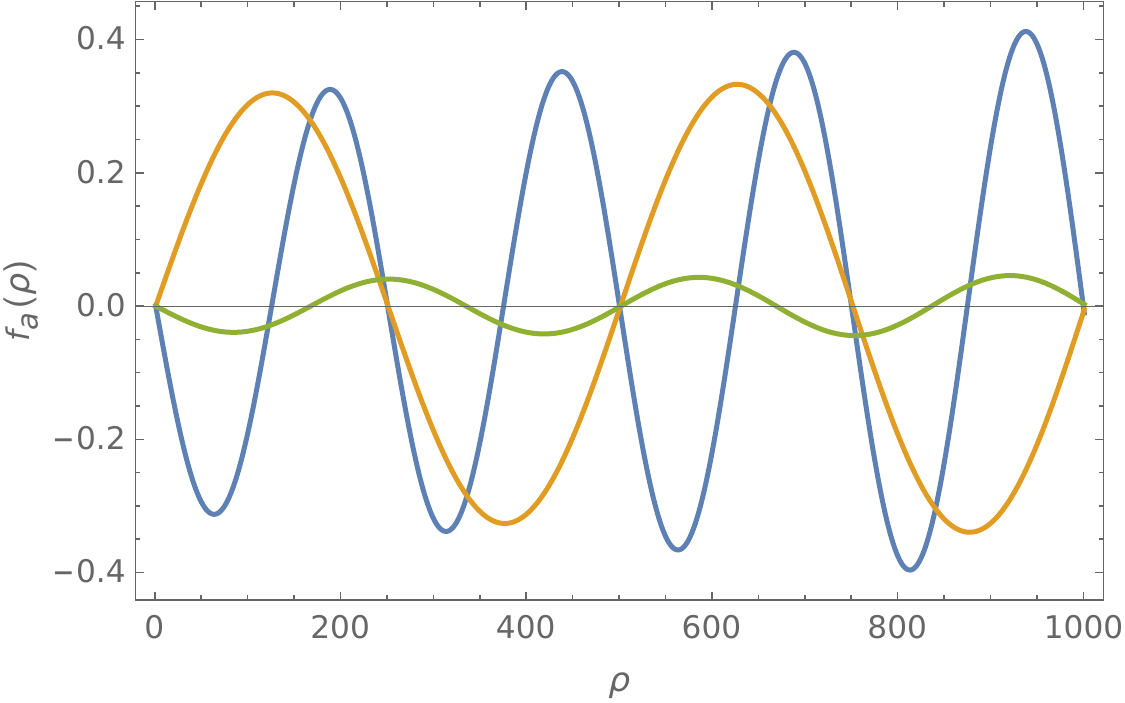}
  \\          \includegraphics[width=7cm, height=6cm]{ 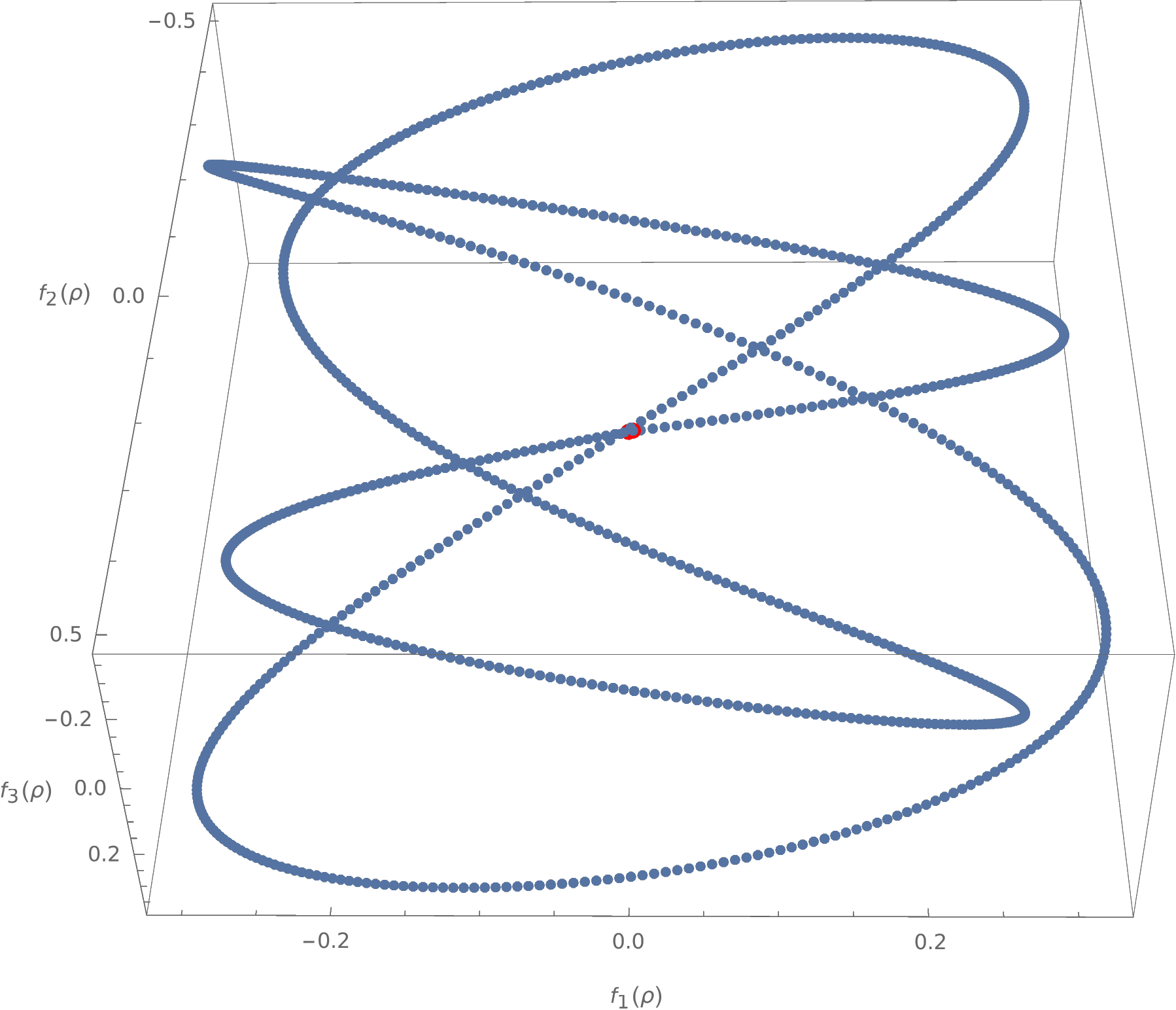}
	\end{center}
	\caption{Numerical solution of (\ref{eq:fNoIsotropico}). Left: the different $f_a$'s are represented in different colors for a unit of the torus lattice. Right: The compatibility between the three $f_a$ solutions is depicted. The red dot in the right figure points out the boundary conditions.}
 	\label{NumericalSolution}
\end{figure}


\noindent
with $x(0) = x_0$, $h = \delta \rho$ y $\delta \rho = 10^{-2}$ . Due to the way the Newton's method is designed, it is desirable to start the process with some initial value as close to zero as possible. But, in the absence of any intuition about where the zero might be in this case, we propose a exhaustive search over a coarse grid of the values of $\omega_i$ (a "free" parameter) and $f_{a}$ that produce a good approach to the fulfillment of the boundary conditions. From these values, Newton method is applied. The Figure \ref{Fig1} show the difference $x(0) - \Phi(x(0),2\pi)$, in each iteration of Newton's method. 

To summarize this section, we obtain a numerical solution for the non-isotropic Q-ball-like soliton. The shape of the individuals $f_a$ are depicted in Figure \ref{NumericalSolution}a. Note that the $Z_a$ solutions must be thought with respect to both  $(\sigma,\rho)$ variables although only the dependence on $\rho$ is depicted. It can be seen  that the amplitudes and the frequencies of the solutions are not constant and they do not correspond to trigonometric functions. Note that they vary over their period. In figure \ref{NumericalSolution}b we can observe  the matching of the three different solutions to the system of equations of motion satisfying all of the three constraints with non-vanishing central charge condition, providing an admissible solution. 


\section{Dynamics of the solutions} In this section we study the dynamics of the analytic solutions previously found under a Lorentz Boost and under a Galilean transformation. Since we are studying the propagation of excitation over the membrane it is not mandatory that their propagation to be relativistic, hence we will analyze the two cases. We will see that the Galilean one is more restrictive. A particularity of our approach is that we work in the LCG and not in a covariant formulation, hence, it is not enough to compute the transformed solutions to guarantee that they still remain being  solutions of the system, hence, it is need to study them with care. In first place we will make general considerations valid for all the solutions considered , alaytic and numerical, that satisfy the ansatz $\partial_{\sigma}f\propto \partial_{\rho} f$.
 
In the case of a Lorentz transformation of a generic family of solutions, we define a group velocity ${\bf{v}}=(v_\sigma,v_\rho)$ with
$v^2=\vert {\bf{v}}\vert^2=v_\sigma^2+v_\rho^2$. The worldvolume coordinate transformations imply 
\begin{equation}
\begin{aligned}
   \sigma\prime&= -\gamma v_\sigma\tau+\sigma+\left(\frac{v_\sigma}{v}\right)^2(\gamma-1)\sigma+\frac{v_\sigma v_\rho}{v^2}(\gamma-1)\rho \,,
   \\
   \rho\prime&= -\gamma v_\rho\tau+\rho+\left(\frac{v_\rho}{v}\right)^2(\gamma-1)\rho+\frac{v_\rho v_\sigma}{v^2}(\gamma-1)\sigma \,,
      \\
   \tau\prime&= -\gamma  \frac{v_\sigma}{c^2} \sigma-\gamma\frac{v_\rho}{c^2}  \rho  +\gamma \tau \,,
\end{aligned}
\end{equation}
with
$\gamma=\frac{1}{\sqrt{1-\frac{v^2}{c^2}}}$ .
The function $Z=f(\sigma,\rho)e^{i\omega\tau}$ becomes modified under the above transformation as
\begin{equation}
\begin{aligned}
Z^\prime(\sigma\prime,\rho\prime,\tau\prime)&=
f(k'\sigma+l'\rho+\omega'\tau) e^{i(\alpha \sigma+\Lambda \rho+\eta \tau)}\,,
\end{aligned}
\end{equation}
where
\begin{equation}\label{eq:relationDinamicaLorentz}
    \begin{aligned}
   k'&:=  k(1+\left(\frac{v_\sigma}{v}\right)^2(\gamma-1) ) +l\frac{v_\rho v_\sigma}{v^2}(\gamma-1)\,,
   \\
   l'&:= l(1 +\left(\frac{v_\rho}{v}\right)^2(\gamma-1) ) +k\frac{v_\rho v_\sigma}{v^2}(\gamma-1)\,,
   \\
   \omega'&:= -\gamma\left(
-k v_\sigma+lv_\rho
\right)\,,
   \\
   \alpha &:=- \gamma  \omega\frac{v_\sigma}{c^2} 
   ,\quad
   \Lambda:= -\gamma \omega \frac{v_\rho}{c^2}
   ,\quad
   \eta:= \gamma \omega \,.
    \end{aligned}
\end{equation}
It describes a travelling soliton. Clearly the transformed solution acquires a spinning Q-ball ansatz when $f$ does not depend on time, i.e.
\begin{equation}
 \omega^{'}=0=  k v_\sigma+lv_\rho,
\end{equation}
which imposes a relation between the two components of the group velocity vector. 
As we have already explained, a transformed solution is not automatically a solution to the equations of motion. On one side,  the functions $f$ must be periodic on the M2-brane worldvolume since $Z$ and $A_r$ are single-valued functions. Also, in order to guarantee that the transformed solutions remain in the set of allowed solutions we restrict ourselves to impose them to be in the family of solutions analyzed in section 3.
The first condition implies that the Fourier modes $(k^\prime, l^\prime)$ must be integers\footnote{In these examples we have fixed the eigenvalue $\lambda$ to one. For arbitrary $\lambda$ the periodicity condition is satisfied imposing $\lambda k, \lambda l\in \mathbb{Z}.$} and requires the following necessary conditions to be satisfied:

\begin{equation}\label{condicion-necesaria}
\begin{aligned}
\left(\frac{v_\sigma}{v}\right)^2(\gamma-1)  \in& Z,\\
\left(\frac{v_\rho}{v}\right)^2(\gamma-1)  \in & Z,\\
\frac{v_\rho v_\sigma}{v^2}(\gamma-1) \in & Z.
\end{aligned}
\end{equation}

They restrict enormously the possible values of the allowed gamma-factor and group velocities. In order to illustrate it let us consider the following example: assume that $v^2/c^2=\epsilon^2= P/Q$ with $P<Q$  and $ v_\sigma=1/N v$, $ v_\rho=1/M v$ with $P,Q,N,M\in Z$, this would imply that the following condition is satisfied for example by requiring that $\epsilon^2=\frac{n^2-1}{n^2}$ which leads to $\gamma=n\in Z-\{0\}$ and if $n-1=M^2N^2$ and  then the group velocities becomes $v_\sigma=\frac{M}{N}v_\rho$ and the necessary conditions are automatically satisfied. Hence, the Lorentz transformed Fourier frequencies are \begin{equation}\label{k transformado}
k^\prime=k(1+N^2)+lMN,\quad l^\prime=l(1+M^2)+kMN.
\end{equation}
In the example considered, one can see that the allowed $\gamma$ coefficients are discrete and rapidly grows towards the speed of light as $n$ grows. It implies that for this choice of the $v/c$ factor, that the group velocity of the propagating excitations  is close to the speed of light, though it does not exclude the existence of other solutions with less group velocity.

These are necessary condition but they do not automatically guarantee the existence of solutions. In the case of the family of solutions verifying (\ref{FamiliaGeneral}) 
and (\ref{FamiliaGeneral-A})  we find that those whose transformed Lorentz $Z$ function must also verify that
\begin{equation}\label{fase lorentz}
    \alpha=k^\prime,\quad \Lambda=l^\prime,\quad \eta=\omega^\prime \,,
\end{equation}
to remain as solutions of the M2-brane with worldvolume fluxes EOM.
See that only $(\alpha, \Lambda)$ impose requirements on 
$$(k^\prime,l^\prime)=\omega\left(\frac{n+1}{n}\right)(\frac{M^2}{v_{\sigma}},\frac{N^2}{v_{\rho}}).$$ 
In the first family case considered in section 3, for the breathing modes $\omega$ given by (\ref{frecuencia omega}) and $k/l=j$, it is possible to obtain a solution by assuming $\omega=Unv_{\sigma}=Tnv_{\rho}$ with $U,T\in Z$. For example there exists a solution for $M=N=2$, $T=5t$, $U=5\frac{M}{N}t$ with $m_1=0, m_2\ne n_2$ and $m_2,n_2,n_1>0$ and $\frac{R_1^2}{R_2^2}=\left(\frac{m_2-jn_2}{jn_2}\right)\left(\frac{jn_2-m_2(1-U)}{jn_1}\right)>0$ with $t< 0$. The transformed Fourier modes become shifted by two different discrete values.

In the case when the $A_r$ real scalar field is included, the gauge field also gets transformed. For the ansatz previously previously considered, 
 with $Z(\sigma,\rho)= f(\sigma,\rho) \ e^{i(\lambda(l \rho+k \sigma)+\tau \omega)}$ its Lorentz transformation corresponds to 
\begin{equation}
f^\prime_a(\sigma^{\prime},\rho^{\prime})= r_a \left(\operatorname{cos}[\lambda(l \rho+k \sigma)] + \operatorname{sin}[\lambda(l \rho+k \sigma)]\right)\,, 
\end{equation}
with the transformed frequency given by \begin{eqnarray}
  \omega'^2&=&\lambda^2 R_{r}^2 \left(k' m_{r}-l' n_{r}\right)^2+\lambda^2 R_{s}^2 \left(k'm_{s}-l' n_{s}\right)^2 \nonumber \\
&=& \lambda^2 \sum_r R_{r}^2 \left(\frac{v_\sigma}{c} m_{r}-\frac{v_\rho}{c} n_{r}\right)^2 \,.
\end{eqnarray}
 See that due to the  $A_r$ EOM gives a frequency 
\begin{eqnarray}
a_r \Omega_r^2= a_r \omega'^2 
&=&\lambda'^2 \left( a_r\left(k' m_{s}-l' n_{s}\right)^2 R_{s}^2 
+
\right.
\\ 
&&
-a_{s} \left( k' m_{r}- l' n_{r}\right) \left(k' m_{s}-l' n_{s}\right) R_{r}R_{s}\Big)\, \nonumber,
\end{eqnarray}

\noindent with the amplitude of $A_r$ given by

\begin{equation}
\begin{aligned}
a_r  =& \frac{\lambda^2}{c^4} ( a_r\left(v_\sigma m_{s}-v_\rho n_{s}\right)^2 R_{s}^2 + \\ 
&-{a_{s}} \left( v_\sigma m_{r}- v_\rho n_{r}\right) \left(v_\sigma m_{s}-v_\rho n_{s}\right) R_{r}R_{s} ) \,,   
\end{aligned} 
\end{equation}

\noindent and verifying the APD transformed constraint. 
\begin{equation}
\sum_r a_{r}\left(k' m_{r}-l' n_{r}\right) {R}_{r}=\sum_r a_{r}\left(v_\sigma m_{r}-v_\rho n_{r}\right) {R}_{r}= 0 \,.
\end{equation}
The previous equations and the ansatz, impose that the frequencies of the Q-ball-like breathing modes and the temporal dependence of $A_r$ to be equal, $\omega^\prime =\Omega^\prime=\Omega_r^\prime$ with $r=1,2, r\ne s$.
 Again,  it acquires a spinning Q-ball ansatz when a relation between the components of the group vector velocity is imposed,
\begin{equation}\label{eq:ConGalileoQball}
     k v_\sigma+lv_\rho=0,
\end{equation}

Let us remind that these solitonic solutions are defined over the M2-brane worldvolume and describe the traveling excitations over the M2-brane with worldvolume fluxes. The center of mass of the M2-brane at this stage propagate as a free particle and its mode is decoupled from the excitations ones \cite{mpgm10Monodromy},
 (as it should happen in case of a particle interpretation). From that point of view, there is not any prerequisite that oblige the excitations to be relativistic, they can behave as classical ones.  
Galileo transformations
are obtained  for $\gamma=1$, satisfying, 
\begin{equation}
\begin{aligned}
   \sigma\prime &= \sigma
   -\gamma v_\sigma\tau, \quad
   \rho\prime &= \rho-\gamma v_\rho\tau,\quad
   \tau\prime &=  \tau \,,
\end{aligned}
\end{equation}
implies a modification in the Q-ball expression that acquires a temporal dependence as a traveling wave, 
\begin{equation}
\begin{aligned}
Z^\prime(\sigma\prime,\rho\prime,\tau\prime)
\equiv
f(k\sigma+l\rho+\omega'\tau) e^{i\omega \tau} \,,
\end{aligned}
\end{equation}
where
\begin{equation}\label{eqn616}
    \begin{aligned}
   \omega'&:= -\left( 
 k v_\sigma+lv_\rho
\right)\,,
    \end{aligned}
\end{equation}

In order to keep the solution stationary inside the Q-ball ansatz it is required that $\omega^\prime=0$, then the same relation between the group speeds hold
\begin{equation}
     k v_\sigma+lv_\rho=0\,.
\end{equation}

 See, that this requirement, needs to be fulfilled for every single function $f$ that satisfies the equations.  Hence, it holds not only of the analytic family of solutions found in section 3 but for the entire range of solutions analytically or numerically modeled. Indeed, from the numerical point of view, the Galilean transformation can be directly imposed and it does not change the picture of the Q-ball solution already obtained. It can be straightforwardly verified that the dynamics of the solutions preserve the the topological central charge condition in all cases, as expected.



\section{Topological and non-topological charges of the M2-brane}
The supermembrane with worldvolume fluxes contains a topological magnetic monopole associated with the topological constraint central charge. We have also seen that some of the solutions that also admits a non-topological Q-ball-like  in the non-compact variables. A natural check is to study the dependence of the breathing frequency with the energy. 

The Q-ball-like solutions have an associated $U(1)$ Noether charge defined as 
\begin{equation}\label{charge_Q-ball ansatz}
Q_a=\omega_a\int d^2\sigma f^2_a(\sigma,\rho).
\end{equation}
in terms of its density charge $q_a$ can be expressed,
\begin{equation}
Q_a=\int d\sigma d\rho q_a(\sigma,\rho), \quad\textrm{with}\quad q_a(\sigma,\rho)=\omega_af_a^2(\sigma,\rho) \,.
\end{equation}
In order to simplify the analysis at the maximum, to see th enon-vanishing contribution we will assume 
the Hamiltonian (\ref{hamiltonianirred}), restricted to the Q-ball-like ansatz, for the isotropic case with the $A_r$ field constant. Then, it becomes, \begin{equation}
H = T^{-2/3}\int_{\Sigma}d^2 \sigma \sqrt{W}
\frac{1}{2}[P_a^2+\frac{T^2}{2}(D_r f_a)^2] \,.
\label{H-charge_1}
\end{equation}

For the family of solutions considered in this paper  (analytical and numerical) satisfying (\ref{FamiliaGeneral}), $\partial_\rho f_a=j_a^{-1}\partial_\sigma f_a$, the quadratic contributions become
\begin{equation}
\sum_r(D_rf_a)^2=\frac{1}{4 \omega_a q_a}\sum_r R_r^2(m_rj_a^{-1}-n_r)^2(\partial_\sigma q_a)^2 \,.
\label{charge3} 
\end{equation}
Substituting  the momentum associated with the $Z$ coordinates in terms of the Noether charge, then the Hamiltonian can be bounded by below by the  following terms
\begin{equation}
H\ge \frac{3}{2}T^{2/3}Q_a\omega_a+\frac{T^{4/3}}{2}\int_{\Sigma}d^2\sigma\sqrt{W}\sum_r(D_rf_a)^2 \,.
\label{H-charge}
\end{equation}
For the particular case of solutions of the family, the Hamiltonian lower bound expressed in terms of the Noether density charge gets simplified to

\begin{equation}
\begin{aligned}
T^{2/3}H &\ge T^{4/3}\frac{3}{2}Q_a\omega+ \\
& +\frac{T^2}{8\omega_a}\sum_r R_r^2(m_r-j_a^{-1}n_r)^2\int_{\Sigma}d^2\sigma\sqrt{W}\frac{(\partial_\sigma q_a)^2}{q_a}\,.  
\end{aligned}
\end{equation}

The mass operator for the M2-brane with fluxes \citep{SpinningSolutionsBosonicM2-brane2022},  is bounded by below by the contribution of the topological charges: the central charge (or equivalently  $C_-$ flux associated with the presence of a monopole), the $C_+$ flux contribution and the Kaluza Klein charges. The non-topological ones associated with the presence of Q-ball-like solutions and the interaction terms between both types of solitons contained in the mass term  $(D_rf_a)^2$
\begin{equation}
\begin{aligned}
\mathcal{M}^2 &\geq Q_{topological}^2
+
T^{4/3}\frac{3}{2}Q_a\omega+ 
\\
& + \frac{T^{2}}{8\omega_a}\sum_r R_r^2(m_r-j_a^{-1}n_r)^2\int_{\Sigma}d^2\sigma\sqrt{W}\frac{(\partial_\sigma q_a)^2}{q_a}\,,
\end{aligned}
\end{equation}
where 
\begin{equation}
\begin{aligned}
Q_{topological}^2=& Q_{fluxes}^2+Q_{KK}^2
\\
=& \int_{\Sigma}C_- 
+ \int_{\Sigma}C_+ 
+\left(\frac{m|q \tilde\tau-p|}{R [I_{m}\tilde\tau]}\right) ^2\,,
\end{aligned} 
\end{equation}
 with 
\begin{equation}
\int_{\Sigma}C_-= 4 \pi^2 R_1R_2 T^2n^2 \,,
\end{equation}
the units of flux, corresponding to the value of the central charge (\ref{central charge}) as originally shown in \cite{mpgm10Monodromy, mpdm2019:flux}. See, that the interaction term contains the Noether charge density of the noncompact sector and the winding numbers required to satisfy the central charge condition that implies the presence of the magnetic monopole. The interaction term is due to the presence of the worldvolume fluxes. Due to it, (responsible for the change in the spectral properties), the characteristic relation between the energy and the frequency of the breathing modes in a standard Q-ball, acquires an extra term associated with the non-vanishing mass terms that add a new dependence on the frequency, that leads a relation between the Energy, the Noether charge, the topological winding numbers and the density Noether charge relations. For $T=1$ it is,

\begin{equation}\label{relacion}
\frac{\partial E}{\partial\omega}=\frac{3}{2}\sum_aQ_a+\frac{1}{8\omega_a^2}\Theta\int_{\Sigma}d^2\sigma\sqrt{W}\frac{(\partial_\sigma q_a)^2}{q_a} \,,
\end{equation}
with $\Theta=\sum_r R_r^2(m_r-un_r)^2$.
In this analysis $f_a$ is an arbitrary function of the family. This type of dependence has been also found in the recent paper \cite{MappinggaugedQ-balls} in which they discuss Q-balls beyond thin-wall limit.

If now we restrict to  the analytic family of solutions considered in section 3, we can observe that for $j_a=1$ we can integrate the function and obtain the standard dependence for a Q-ball $\frac{\partial E}{\partial\omega}=\frac{3}{2}\sum_aQ_a$ shifted by a constant term.

\paragraph{Stability of the Solutions}
An important point that has not been addressed so far is the stability of the solutions found. The monopole charge induced by the fluxes acting on the Hamiltonian through the non-vanishing covariant derivative "mass term", associated to the monopole and Q-ball interaction term, protects the worldvolume of the membrane from fission at the classical level, since there are no string-like solutions with vanishing energy that can split and merge the membrane without energy cost when the area of the worldvolume is conserved \cite{gmr}. It is well known that this instability classically occurs over generically for the membrane, \cite{dwhn}. In that scenario, a Q-ball excitation would not make sense, if defined on an object that can split and merge without any energy cost as is the case for the worldvolume of the flux-free supermembrane, i.e. vanishing central charge. Therefore, the flux condition is a necessary condition for the stability of the Q-ball against fission.

On the other hand, we expect that the fact that the interaction term  (Covariant derivative mass terms) is non-vanishing,  will provide a sufficient condition for the stability against fission of the Q-ball-like excitations that propagates on the membrane.  Indeed, this is the case for the analytic solutions of the string Q-ball-like excitations considered in this paper, as it can be straightforwardly checked. Given a fixed time t, those Q-ball-like solutions require an addition of energy to become splitted. Furthermore, since the functions considered are periodic, the stability in the propagation of those solutions is also guaranteed. We expect to holfd this property generically.  However, to ensure it, this needs to be proved for arbitrary functions. This question remains unanswered so far and is beyond the scope of this paper.

\section{Superposition of isotropic solutions: Vectorization}
An important aspect of solitonic solutions is the fact that on very general grounds, solitons cannot be superposed linearly. In fact, this is a characteristic property, however, there are some exceptions, that for a restricted range of frequencies, admit linear superpositions like it is the case for optical solitons,  \cite{SuperpositionsBrightanddarksolitons,LinearSuperpositionPonomarenko2002,ZHOU20171697:LinearSuperpositionPrinciple}. We show in this section that this is also the case of the Q-ball-like solutions analyzed in section 3. We will see that in the absence of a symplectic gauge field, we are able to obtain multi-soliton solutions that naively seems to corresponds to a linear superposition but only when restricted to a very specific discrete set of frequencies given by the monopole contribution. In order to understand it better we will do it in several steps. 
\subsection{Fourier series solution} Although we are interested in Q-ball-like solutions, we will start by analyzing the behaviour of the EOM of the M2-branes with worldvolume fluxes for the case of a superposition of the complexified analytic solutions found previously in section 3. We can observe that they correspond to a Fourier serie. Since this is a sum, with matrix entries, it is not automatically guaranteed that the superposition of solutions will be also a solution due to the non-linearities of the differential equations. Indeed we are interested in finding the conditions under which this may happen. 
Let us consider $f_a(\sigma,\rho)$ represented by a Fourier series:
\begin{equation}
f_a^{\mathbb{C}}= \sum_{k,l=0}^{\infty}  a_{k l} f^{kl}_a(\sigma,\rho) \,.
\label{Fourier-trig}
\end{equation}
The $kl$-th element of the orthogonal basis is given by:
\begin{equation} 
f^{kl}_a(\sigma,\rho) = \cos(k \sigma + l \rho) + i \sin(k \sigma + l \rho) \,.
\end{equation}

It is easy to verify that, with a proper normalization:

\begin{equation} 
\int_{0}^{2 \pi} \int_{0}^{2 \pi} f^{k
l}_a(\sigma,\rho)~ f^{uv}_a(\sigma,\rho)~ d\sigma~ d\rho =
 \begin{cases} 
0 & uv \neq kl \\
1 & uv = kl
\end{cases}
\end{equation}
%
with,
\begin{equation}
a_{nm} = \int_{0}^{2\pi} \int_{0}^{2\pi} f_a(\sigma,\rho)~  f^{kl}_a(\sigma,\rho)~ d\sigma d\rho \,.
\end{equation}

Imposing smoothness on $f$ and differentiability  term-by-term, see for example section $3.4$ in \cite{Haberman}. The series  (\ref{Fourier-trig}) can be written as:

\begin{equation}
\label{Fourier-trig-1}
f_a^{\mathbb{C}}= \sum_{ \xi_u=0}^{\infty} a_{\xi_u}e^{i ~\xi_u. u} \,,
\end{equation}
%
and derivatives
{\footnotesize
\begin{equation}
\begin{aligned}
\partial_{u}  f_{a}(u)  &=  \sum_{ \xi_u=0}^{\infty}a_{\xi_u} i~\xi_u~ e^{i ~\xi_u. u}, \ \ 
\partial_{u}^2 f_{a}(\sigma,\rho),  = \\
&= -\sum_{\xi_u=0}^{\infty} a_{\xi_u}\xi_u^2~ e^{i \xi_u u}, \ \ 
\partial_{u v}^2  f_{a}(\xi_u)  =  -\sum_{
 \xi_u=0}^{\infty}  a_{\xi_u}\xi_u\xi_v~ e^{i ~\xi_u u},    
\end{aligned}
\end{equation}
}

with $ u \not= v$. We have defined the vector $u=(\sigma,\rho)$ labeling the spatial coordinates and the vector $\xi_u=(k,l)$ labeling the Fourier modes of the expansion, and $a_{xi_u}$ the constant amplitude of $f_a^{\mathbb{C}}$. 
\noindent

\begin{figure}[ht]
	\begin{center}
		\includegraphics[width=0.23\textwidth]
  { 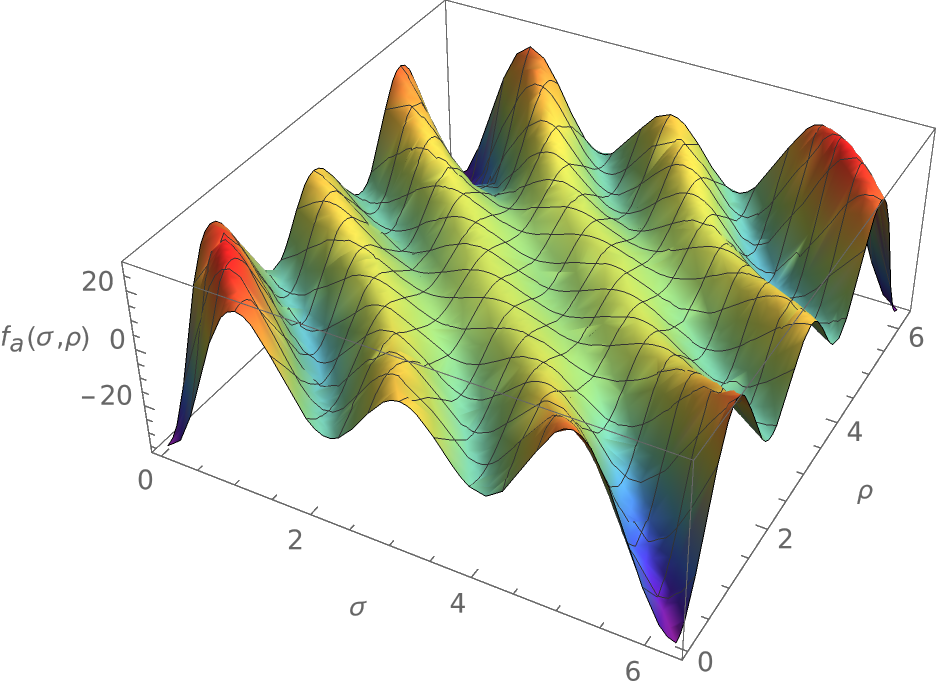}	\includegraphics[width=0.23\textwidth]
{ 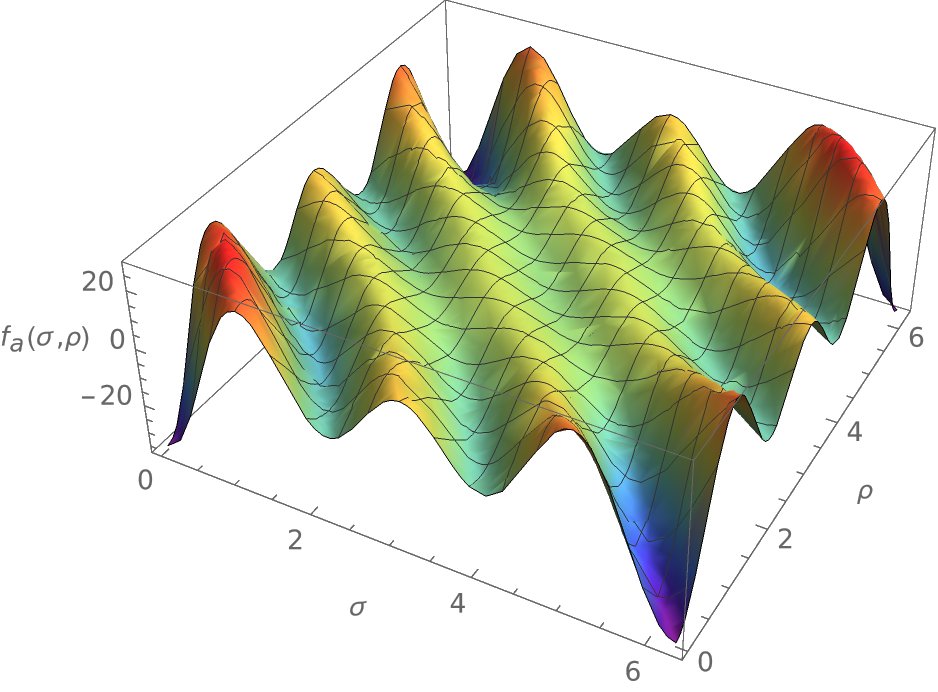}
	\end{center}
	\caption{The Figure of  left represents the analytic derivative behaviour and the Figure of the right  "{\it inner product version}" of the derivative in the case of $\partial^2_{\sigma \rho} f_{a}(\sigma,\rho)$. As we can observe both graphics are indistinguibles.}
	\label{Fig:Analitic-Inner}
\end{figure}

To implement the expansion numerically, we use a truncated version of the series (\ref{Fourier-trig}) and define, the vectors:

\begin{equation}
\begin{aligned}
{\bf k} & = \left(0, 1, 2, 3, \cdots N ,  1, 2, 3, \cdots, N, \cdots \right) \,, 
\\
{\bf l}  & =   \left(0,  0, 0, 0, \cdots 1 ,  1, 1, 1, \cdots, N, N,N,\cdots \right), 
\\
{\bf k}^2 & = \left(0,  1, 4, 9, \cdots, k^2, \cdots N^2\right), 
\\
{\bf l}^2  & =  \left(0,   0, 0, 0,  \cdots, l^2, \cdots N^2\right), 
\\
{\bf kl} & =   \left(0, 0, 0,0; 0,1,2,3,\dots; 0,2,4,6,\cdots, k l, \cdots, \right), 
\\
{\bf f}_{a}(\sigma,\rho) & = (  a_{11} ~f_{a}^{11} (\sigma,\rho),  a_{12} ~f_{a}^{12}(\sigma,\rho), \cdots, 
\\
& \cdots,  a_{1N} ~f_{a}^{1N}(\sigma,\rho), a_{21} ~f_{a}^{21} \cdots  a_{NN} ~f_{a}^{NN} ...),  
\\
\mathbf{\omega}^{2} & = ( \omega_{11}, \omega_{12}, \omega_{13}, \omega_{kl}, \cdots, \omega_{NN} ).
\end{aligned}
\end{equation}
See also that ${\bf l}$ must be defined in a different way than the others indices.
With the above notation \footnote{There, the vectors $k$ and $l$ are a concatenation of $N$ vectors of integers from $0$ to $N$ and ${\bf f}_{a}(\sigma,\rho)$ is a flattened version of the matrix $f_{a}^{kl}$.} defined and considering only $Sine$ or $Cosine$ series, we can write:
\begin{eqnarray}
\partial_{\sigma}  f_{a}(\sigma,\rho) &=&  i~  {\bf k} \cdot \textbf{f}_a(\sigma,\rho), \nonumber \quad
\partial_{\rho} f_{a}(\sigma,\rho)  =   i~  {\bf l} \cdot \textbf{f}(\sigma,\rho), \nonumber
\\ 
\partial_{\sigma} ^2 f_{a}(\sigma,\rho), & = & - {\bf k}^2 \cdot \textbf{f}(\sigma,\rho), \nonumber \quad
\partial_{\rho}^2 f_{a}(\sigma,\rho)  =  - {\bf l}^2  \cdot  \textbf{f}(\sigma,\rho), \nonumber \\ 
\partial^2_{\sigma \rho} f_{a}(\sigma,\rho)  &=&  - {\bf kl} \cdot   \textbf{f}(\sigma,\rho),
\label{Der-Prod}
\end{eqnarray}

where, $(\cdot)$ denotes a inner product.
See that obviously this is not a general property but a property of the Lie bracket and the base chosen that we know that satisfies the equations in the case of a single solution.
The Figure (\ref{Fig:Analitic-Inner})  compares in a graphic fashion, the "{\it analytic}" derivative and "{\it inner product version}" of the derivative in the case of $\partial^2_{\sigma \rho} f_{a}(\sigma,\rho)$. See that the agreement is very good.

The system that we intend to solve corresponds to the EOM  (\ref{EOM_Zc-Qball}) and (\ref{EOMForArZiagualfg-Qball}) with $A_r=constant$
\begin{equation}
\label{System1}
-C_{1}\left(\partial_{\sigma} ^2 f_{a}\right)+ 2~ C_{2}\left(\partial^2_{\sigma \rho} f_{a}\right)-C_{3} \partial_{\rho}^2 f_{a}=\omega^{2}f_{a} ,
\end{equation}
\begin{equation}
 \label{System2}
\left(m_{r} \partial_{\sigma}^{2} f_{a}-n_{r} \partial^2_{\sigma \rho} f_{a}\right) \partial_\rho f_{a}
-
\left(m_{r} \partial^2_{\rho \sigma} f_{a}-n_{r} \partial_{\rho}^{2} f_{a} \right)\partial_{\sigma} f_{a} = 0 \,,   
\end{equation}
\newline
where
$
C_1 = m_9^2 R_9^2 + m_{10}^2 R_{10}^2$,  
$C_2 = m_{9} n_{9} R_{9}^2 + n_{10} m_{10} R_{10}^2$, 
$C_3 =  n_9^2 R_9^2 + n_{10}^2 R_{10}^2,$ 
with $r= 9, 10$,  $n_r, m_r$, the winding numbers, $R_r$ the torus radii and $\omega \in \mathbb{R}$. The APD constraint is identically verified. It is subject to the central charge condition (\ref{central charge}).

We impose now the ansatz of the family (\ref{FamiliaGeneral}) for the isotropic case. 
The equation (\ref{System1}) written in the \textit{inner product version} results in:
\begin{eqnarray}
\left( C_1~ {\bf k}^2 - 2~  C_{2}~ {\bf kl}  + C_{3} {\bf l}^2 \right) \cdot \textbf{f}(\sigma,\rho) & = & \mathbf{\omega}^{2} \cdot \textbf{f}(\sigma,\rho),
\end{eqnarray}

\noindent
where each component, it can be expressed directly in terms of their breathing modes: 

\begin{equation}
C_1~ k^2 - 2~  C_{2}~ k l  + C_{3}~ l^2  =\omega_{kl}^{2}\,.
\end{equation}

The second equation (\ref{System2}), since it contains the nonlinearities is more involved, but still can be expressed in this simple algebraic way:
\begin{equation}
 \label{NPDE-Fourier}
 \begin{aligned}
-[\left( m_{r} {\bf k}^2 - n_{r} {\bf kl} \right) \cdot \mathbf{f}(\sigma,\rho)] \times[~ {\bf l} \cdot \textbf{f}(\sigma,\rho)] &+ \\
+[\left(m_{r} {\bf kl} - n_{r} {\bf l}^2 \right)\cdot \mathbf{f}(\sigma,\rho)] \times[~ {\bf k} \cdot \textbf{f}(\sigma,\rho)] & = 0.  \end{aligned}
\end{equation}

Here $\times$ represents the product between scalars or scalar and vectors.
In order to solve this equation, let us define the vectors:
\begin{equation}
{\boldsymbol \alpha}_r =  m_{r} {\bf k}^2 - n_{r} {\bf kl}, \quad
{\boldsymbol \beta}_r = m_{r} {\bf kl} - n_{r} {\bf l}^2 \,,
\end{equation}

\noindent
with components $\alpha_{m}$ and $\beta_{m}$, respectively, and being ${\bf f}_{mn} $ represents the matrix elements of ${\bf f}$. The equation 
(\ref{NPDE-Fourier}) expressed in terms of $\boldsymbol \alpha, \boldsymbol \beta$ becomes
\begin{equation}
- (\boldsymbol \alpha_r \cdot \mathbf{f}(\sigma,\rho))\times ({\bf l} \cdot \textbf{f}(\sigma,\rho)) +
(\boldsymbol \beta_r \cdot \mathbf{f}(\sigma,\rho))\times( {\bf k} \cdot \textbf{f}(\sigma,\rho)) = 0 \,.
\label{Real-problem}
\end{equation}

Before continuing, it is interesting to note that if the series (\ref{Fourier-trig}) has only one term, the equation (\ref{System2}) is identically satisfied, since,
\begin{eqnarray}
n_{r}~ l^2 ~k - m_{r}~ l~k^2  + m_{r}~ l ~ k^2 - n_{r}~k l^2  &=& 0 \,,
\label{Observation}
\end{eqnarray}
which agrees with the results of Section 3.
therefore, The eigenvalues and eigenvectors obtained between the analytical and numerical methods are very similar to those found in \cite{ SpinningSolutionsBosonicM2-brane2022} although in that case they were  numerical solutions to the  approximate  system, (i.e. without considering the restriction of (\ref{System2})). This means that the simplest solutions obtained is in the kernel of the monopole restriction. 
\subsection{Solving an algebraic problem}
In order to find a larger set of solutions, we have transformed our initial problem into another one: find a $\bf f$ that fulfill (\ref{Real-problem}). Since (\ref{Real-problem}) can be written as,
\begin{equation}
\left[ -\boldsymbol \alpha_r \cdot \mathbf{f}(\sigma,\rho) \times~ {\bf l}  +
\boldsymbol \beta_r \cdot \mathbf{f}(\sigma,\rho) \times~ {\bf k} \right ] \cdot \textbf{f}(\sigma,\rho) = 0.
\end{equation}

\noindent
There are two alternatives: one can linearly transform $\bf f$ (selecting the set of coefficients $a_{mn}$), so that obtain a vector
\begin{equation}
 -\boldsymbol \alpha_r \cdot \mathbf{f}(\sigma,\rho) \times~ {\bf l}  +
\boldsymbol \beta_r \cdot \mathbf{f}(\sigma,\rho) \times~ {\bf k},
\end{equation}	
\noindent
orthogonal to $\bf f$ (this seems to be an interesting, but complicated problem), or more easily, we can suppose that ${\bf f} \neq 0$, and find ${\bf f}$ (the set of coefficients $a_{mn}$ that satisfy the equation), that is:
\begin{equation}\label{vectorizacion}
\boldsymbol \alpha_r \cdot \mathbf{f}(\sigma,\rho) \times~ {\bf l}  =
\boldsymbol \beta_r \cdot \mathbf{f}(\sigma,\rho) \times~ {\bf k}  \,.
\end{equation}

\noindent
In order to find a solution to ((\ref{System2})), we denote the position of an element by $l_j$, in the corresponding vector, say for example ${\bf{l}}$. Then the solution with the allowed values is found when  $k_j=l_j$ for $j=i+(i-1)N$ being $N$ the order of the truncation and $i$ the position inside it, $i=1,\dots,N$. The coefficient values in those positions are given by $l_j=k_j=i-1$. The solution then arises for the subfamily of functions $f^{\mathbb{C}}$ with $a_{k_jl_j}=1$  and zero otherwise. It corresponds to the diagonal terms of $\textbf{f}$ in a matrix representation. 

\noindent
Hence, we obtain that the superposition of arbitrary functions with equal Fourier modes is solution to the EOM (\ref{System1}) and (\ref{System2}) whenever those modes are equal. This condition, constrains the set of solutions of  equation (\ref{System1}), and reflects the nonlinearity in behavior to the equation (\ref{System2}), associated with the presence of a monopole solution constructed in terms of the $\widehat{X}_r$.  

In short, a general solution for the EOM (\ref{System1}) and (\ref{System2}) system with a QBL ansatz and nonvanishing central charge can be written in terms of the real parts of $f^{\mathbb{C}}$. It is straightforward to see that the same result holds for 
\begin{equation}
f_a(\sigma,\rho,w) = \sum _{k=1}^M a_{kk} \left[  \cos (k (\rho  + \sigma)) +  \sin(k( \rho  +  \sigma)  \right] \,.
\label{SolIm}
\end{equation}

The sum  runs over the number of pairs $M$ that fulfill the solution. 
It represents a Q-ball-like multisoliton solution for the isotropic case with a non vanishing central charge.  

In summary, the analytic family of QBL solutions for the isotropic case and constant gauge field and with central charge different from zero admit a linear superposition law but only when the frequencies become strongly constrained to be equal.  We emphasize that this is not true otherwise. All of these case require a central charge different from zero -which we recall it is associated to the presence of a magnetic monopole over the worldvolume \cite{MARTINRestucciaTorrealba1998:StabilityM2Compactified}. This represents the superposition law for this type of solutions.

 In the following we illustrate an example of superposition of solutions.


Here, it can be seen how the proposed solution complies with the equation (\ref{System1}-\ref{System2}), with $\{k_j,l_j\}_{j=0}^{M} = \{\{0, 0\}, \{1, 1\}, \{2, 2\}, \{3, 3\}, \{4, 4\}\} $  and  $m_9=9$, $m_{10}=2$, $n_9=9$, $n_{10}=8$, $R_9=1$ and $R_{10}=11$ with non-vanishing central charge.



\begin{figure}[ht]
	\begin{center}
	\includegraphics[width=7cm, height=5cm]{ 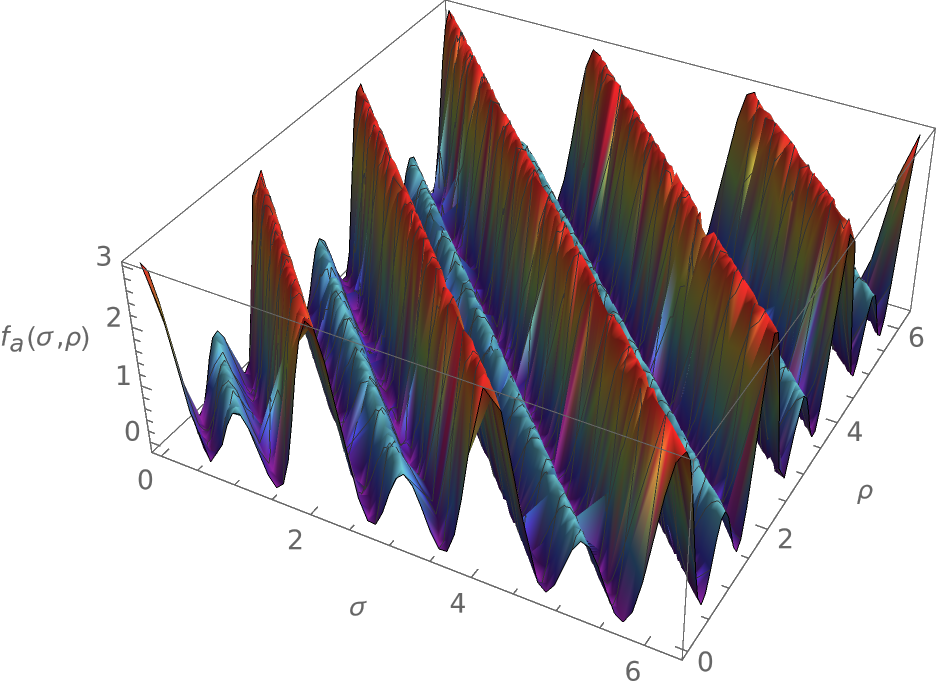}
	\end{center}
	\caption{Analytic solution of (\ref{System1})-(\ref{System2})} 
\end{figure}


We observe the following: i) The equation (\ref{Real-problem}), is very interesting since it allows to reduce the order of non-linearity in this type of equations, ii) The condition (\ref{System2}) seems to act as a pruning to the basis (solutions) generated by the eigenvalue (\ref{System1}) ad iii) The linear superposition is not possible on general grounds due to the nonlinearities of the equation (\ref{System2}). However, it can be obtained when the values of the Fourier modes becomes strongly constrained to specific pairs that determine the frequencies due to the nonlinear behaviour of (\ref{System2}).  There are in the literature other examples of nonlinear EPD that admit restricted linear superposition, see \cite{Han2022}
.


\section{Discussion and Conclusions}

In this paper we obtain Q-ball-like solitonic solutions to the dynamics of the bosonic part of a supermembrane compactified on a target space $M_9\times T^2$  with a nontrivial $C_{\pm}$ fluxes. We have focused  in obtaining solutions with a  Q-ball-like ansatz defined on the M2-brane worldvolume. These are solutions to the complex scalar fields $Z_a$ that determine the embedding of the supermembrane on the  non-compact dimensions. We have analyzed them satisfying the ansatz $Z_a=f_a(\sigma,\rho)e^{i\omega\tau}$. The name of Q-ball has been maintained, though the solutions characterize solitons over a torus, -since  the spatial part of the worldvolume of the supermembrane is assumed to be Riemann surface of genus one $\Sigma_1$. The system is in the presence of a symplectic gauge field $\mathbb{A}$ on the M2-brane worldvolume whose components are defined in terms of derivatives of a single-valued real scalar field $A_r$. Both fields $Z_a$ and $A_r$ define the dynamics of the M2-brane with a system of nine coupled nonlinear equations of motion. The dynamics is restricted by two constraints, the area preserving diffeomorphisms (APD) that is the residual first class constraint of the membrane theory, and the flux condition. This last topological condition implies the so-called \textit{central charge condition} \cite{MARTINRestucciaTorrealba1998:StabilityM2Compactified}, which implies a restriction on the M2-brane wrapping numbers and induces  the presence of a monopole charge.

We obtain analytical solutions to the isotropic case $Z_a=Z$  for all $a=1,2,3$, and numerical solutions in the non-isotropic case $Z_a\ne Z_b$.  The analytic solutions are obtained in the presence of a non-vanishing $A_r$ field and attending to the type of time dependence imposed on the $A_r$ fields, they are organized into different families.  All the cases generate a symplectic gauge field $\mathbb{A}=dA$.  In the first case,  the temporal dependence on $A_r$  is linear.  The frequencies characterizing the Q-ball breathing modes are discrete and they are parametrized in terms of  the compactified target space moduli, the nontrivial wrapping of the M2-brane, and the amplitudes of the $A_r$ field. The Area Preserving Diffeomorphisms (APD) constraint is automatically satisfied. We obtain dynamical solutions to the system that satisfy the central charge restriction and hence the worldvolume flux condition.  In the second family, the dependence of $A_r$ with time, $\tau$, is generalized. The associated symplectic field strength depends on all of the variables. Due to the ansatz chosen and the APD constraint, a restriction to the EOM appear that leads to a relation between the dispersion relation of the Q-ball breathing modes $\omega$ and the dispersion relation $\Omega$ associated with the $A_r$ field. Both frequencies remain discrete. The APD constraint impose a restriction in the amplitudes of the $A_r$. For a fixed relation between them, a ratio between the Fourier modes is determined.  The type of solutions considered satisfy the central charge condition for winding charges suitably chosen. We also provide an example of a localized family of solutions propagating over the worldvolume of the membrane. Other examples even more sofisticated of localized solutions can be constructed. 

We also study their dispersion relations. Generically, the solutions exhibit dispersion although for certain combinations of parameters they also admit solutions dispersion-free. Usually, solitons are associated with solutions in which the dispersion effects are compensated by the non-linearities of the equation. In our case we prove that these type of solitonic solutions free of dispersion effects exist, but also a second type of solitons, that in spite of the dispersion effects, and due to the periodicity of the functions on the M2-brane worldvolume remain being solitons. 
Examples of periodic solitons with dispersion effects have been considered in the context of optic solitons \cite{Dispersion-Managed:TURITSYN2012}. The dispersion effects in this context do not mean that the system present any loose in energy of the system. Energy is conserved since it corresponds to the excitations on a single M2-brane that propagates freely in the space. \newline

We find also solutions to the non-isotropic case $Z_a\ne Z_b$ by performing a numerical analysis.  From the viewpoint of numerical computation, we have proposed and implemented strategies to find  approximate solutions of the Q-ball in the isotropic and non-isotropic cases. We have used the isotropic case to compare the numerical method with the results previously obtained in the analytical case. We use a fine-difference scheme is used to solve the eigenvalue problem (\ref{System1}), and then the resulting eigenfunctions are used to represent the solution of (\ref{System2}). This transforms the equation (\ref{System2}) into a system of nonlinear algebraic equations, whose solution can be obtained using methods such as Newton's method. The non-isotropic case for constant $A_r$ introduces the non-linearities of the $Z$ equation (not only due to the $A_r$ EOM). 
In this case, the solution of the system (\ref{eq:fNoIsotropico}), is approximated using a shooting method where the gradient of the solution is approximated using finite differences.\\
In both cases, very simple approximation schemes are used with solutions that present reasonable approximation errors, so we suppose that more sophisticated schemes could produce better solutions even though they have a higher computational cost.
Hence, for the isotropic and non-isotropic case, with a vanishing symplectic gauge field, we are able to obtain numerically a Q-ball soliton solutions to the M2-brane with worldvolume fluxes. 

The dynamics of the analytic families of solutions is also investigated. We consider Lorentz transformations. Since we work in the LCG, the transformed solutions are not automatically solutions of the EOM. In this sense, we have shown some examples of where we have obtained the necessary restrictions to satisfy the EOM.  Since we are considering solutions that represent solitons defined over the worldvolume, -they correspond to excitations on a M2-brane which propagates freely on the space-, it is not necessary that the excitations will be relativistic. Hence, we also study their Galilean transformations. For this case we find, for the examples considered, that the Galilean transformations restrict further the solutions allowed. We find that for the ansatz considered that only those that preserve their Q-ball shape are allowed, i.e. only contain breathing modes but the pulse does not propagate. However, it does not exclude that more general solutions could be obtained by choosing different ansatzs. Each of the dynamical solutions preserve the flux condition. 

The M2-brane mass operator is bounded by below by the topological contribution due to the flux condition and the Kaluza-Klein terms. For the type of solutions considered in this paper, the kinetic term generates a Q-ball contribution, that also contributes to the lower bound. It also appears a new interaction term between the topological and non-topological sectors in the mass term contribution.
The M2-brane with worldvolume fluxes induces a central charge condition that describes a magnetic monopole \cite{MARTINRestucciaTorrealba1998:StabilityM2Compactified}, and at the same time it may also describe a non-topological Q-ball soliton. We show that there are nontrivial interaction terms between these two, through the mass terms of the potential containing the winding charges that generate the monopole charge and the Noether charge. They describe a relation between the energy $E$ and the frequency $\omega$ that contains on top of the standard term characterizing the Q-ball relation, and a new term that represents a correction to the Q-ball relation reflecting the interaction among those terms and the compactified and non compactified sectors. This type of correction, that here has a topological origin, has also appeared in the literature, in a different context,  when going beyond the thin-wall limit \cite{Heeck_2022:ExcitedQ-balls, MappinggaugedQ-balls, Nugaev2020:ReviewNontopologicalSolitons}.  Recently, it has been discovered a new type of solitons which share these two types of properties. They have been named as \textit{monopole-Q-ball} soliton and their Q-ball solutions show an increased stability due to the effect of the topological charge. It would be interesting if both models are related. \newline

In this case the stability of the membrane against fission is guaranted because of the discreteness of the mass operator spectrum of the bosonic membrane that at classical level forbids the existence of flat directions associated to the presence of string-like spikes \textit{without energy cost} that could induce the non-conservation of the topology nor the number of particles. In the present model, the string-like spikes contribute with a positive energy to the spectrum mass operator of the membrane. This is a special property that for the supersymmetric case is not guaranteed for the target space considered unless the nontrivial worldvolume fluxes are present. The interaction term is responsible for that change in the nature of the spectrum  providing more stability to the Q-ball. In the case of the analytical solutions it is easy to prove that also the stability of the QBL excitation is also guaranteed since its split is energetically disfavoured. We believe that this is a general property, however a more general statement for arbitrary functions requires a study beyond the scope of the present paper and are left for a future work. 

We also investigate the superposition law for the analytic family of solutions found. To this end we find a new method that we denote \textit{vectorization}, that allows to convert the non-linear differential equations in terms of algebraic expressions, much simpler to solve. We find that, in spite that the EOM associated with the field $A_r$ is identically satisfied for a single soliton, due to the nonlinearities of the system of equations, this equation has to be taken into account in the superposition of solutions. We obtain a multi-solitonic solution characterized by a linear superposition, but only when the Fourier modes $k,l$ are equal.  This is consistent with the fact that they are solution to the system of nonlinear equations. In the literature it has appeared other optical soliton cases that  under restricted conditions admit linear superpositions. See for example,  
\cite{LinearSuperpositionPonomarenko2002,ZHOU20171697:LinearSuperpositionPrinciple, SuperpositionsBrightanddarksolitons}.

In summary, we have obtained Q-ball-like solitons solutions of the dynamics of the supermembrane propagating on $M_9\times T^2$ once that the flux condition is imposed. See, that the non-linearities  are present even in the case when the  symplectic gauge field vanishes and this is due to the residual equation associated with the presence of multivalued modes that generate the topological magnetic monopole soliton and from those inherited from the quartic potential contribution. \newline

 \acknowledgments 
 MPGM thanks to G. Alvarez, A. Bellorin  and A. Restuccia for helpful conversations. MPGM is partially supported by the PID2021-125700NB-C21 MCI Spanish Grant. R.P. acknowledge to the Doctorado en Física, mención Física Matemática (Ph.D. in Physics) program of the Universidad de Antofagasta, Chile and thank to the projects ANT20992, ANT1956 y ANT1955 of the U. Antofagasta. M.P.G.M., and R.P. want to thank to SEM18-02 project of the U. Antofagasta.

\bibliography{referencesNot-URL}
\end{document}